%% file: 01_main.tex
\documentclass[11pt]{article}
\usepackage[a4paper,margin=2.2cm]{geometry}
\usepackage[T1]{fontenc}
\usepackage[utf8]{inputenc}
\usepackage{mathptmx}
\usepackage{microtype}
\usepackage{graphicx}
\usepackage{booktabs}
\usepackage{longtable}
\usepackage{amsmath,amssymb}
\usepackage{xcolor}
\usepackage[numbers,sort&compress]{natbib}
\usepackage{setspace}
\usepackage[labelfont=bf,font=small,labelsep=period]{caption}
\usepackage[hidelinks]{hyperref}
\usepackage{enumitem}
\graphicspath{{figs/}}
\newcommand{\fbar}{\bar f}
\newcommand{\fmean}{\langle\bar f\rangle}
\newcommand{\Nagents}{N}
\title{\bfseries Warned alike, AI agents avoid the less-crowded road while people take it}
\author{Takahiro Ezaki$^{1,\ast}$, Naoto Imura$^{1}$, Katsuhiro Nishinari$^{1,2}$\\[4pt]
\normalsize $^{1}$Research Center for Advanced Science and Technology, The University of Tokyo,\\ \normalsize 4-6-1 Komaba, Meguro-ku, Tokyo 153-8904, Japan\\
\normalsize $^{2}$Department of Aeronautics and Astronautics, School of Engineering, The University of Tokyo,\\ \normalsize 7-3-1 Hongo, Bunkyo-ku, Tokyo 113-8656, Japan\\
\normalsize $^{\ast}$Corresponding author. Email: tkezaki@g.ecc.u-tokyo.ac.jp}
\date{}
\begin{document}
\maketitle
\begin{abstract}
AI agents built on a few shared models increasingly act for many people. A shared forecast about others can align their choices and change how scarce capacity is allocated. We tested this feedback in a two-road congestion game. Adding one sentence warning that others might follow a routing tip made populations of 50 GPT agents crowd one road while avoiding the nearly empty alternative. Average travel time rose from 64 to 95 min, although any crowded-road agent could have saved 69 min by switching alone. The warning discouraged the very move it predicted. The pattern persisted for 100 rounds. Two other model families shifted the same way without locking onto one road. Twelve all-human groups (240 participants) stayed near balance under numerical reports or the tip and warning. In 24 mixed groups with a further 240 participants, imbalance grew with the share of agents in the registered analysis, while people increasingly took the road the agents avoided. Collective costs stayed below the all-agent reference, but with 15 agents and 5 humans, agent seats averaged 80 min, compared with 44 min for human seats. Shared forecasts can thus sustain collective inefficiency among similar agents. A better group average can also hide an unequal burden. Evaluations of AI agents that share resources should test populations, treat messages as interventions and report who bears the costs.
\end{abstract}


\section*{Introduction}
As AI systems act on behalf of more people, their collective behavior \citep{rahwan2019,brinkmann2023,burton2024} becomes a problem of resource allocation, not just individual accuracy. Many agents built from the same model may interpret a shared message alike, linking otherwise separate decisions \citep{kleinberg2021,bommasani2022}. This matters wherever users compete for limited capacity, from roads and computing services to electricity demand \citep{macfarlane2019,mitzenmacher2001,ramchurn2011}. A forecast about others' choices is especially consequential. Public signals carry disproportionate weight when people coordinate \citep{morris2002}. Once broadcast, a forecast also changes the situation it describes and can defeat its own prediction \citep{merton1936,grunberg1954}. Can such a forecast sustain costly alignment even when agents repeatedly experience its consequences, and what happens to people who share the same resource?

Congestion games make this feedback observable: the value of a choice depends on how many others make it \citep{arthur1994,challet1997}. In a symmetric two-road game, equal use minimizes travel time and no commuter can gain by switching alone \citep{wardrop1952}. Human experiments often approach this benchmark through heterogeneous choices \citep{selten2007} and learning \citep{helbing2005,erev1998}. Shared information can instead synchronize responses \citep{arnott1991} and produce overreaction, oscillation or persistent disequilibrium \citep{benakiva1991,wahle2000,klugl2004}. Experiments and theory show why better information need not improve individual choices \citep{avineri2006} or system performance \citep{arnott1991,benelia2015,acemoglu2018}. This simple environment therefore tests a general tension between anticipating others and maintaining diversity of action.

Language-model populations can develop conventions and collective biases \citep{ashery2025}, yet struggle when success requires different agents to choose different actions \citep{ballestero2026} or to restrain their use of a shared resource \citep{piatti2024}. Their economic choices can resemble human choices while showing less heterogeneity and weaker coordination in repeated interaction \citep{mei2024,chen2023,akata2025}. Algorithmic pricing likewise shows that individually generated decisions can produce collective outcomes sensitive to learning and prompts \citep{calvano2020,fish2024}. Studies of model commuters have examined equilibrium, learning, memory and differences from human play \citep{wangtraveler2025,liu2025abm,liu2026dual,goodyear2025,wangisci2025}, with related dynamics in the El Farol bar problem, where people choose whether to attend a bar that is enjoyable only when not too crowded \citep{arthur1994,takata2025}. What remains unclear is whether a shared forecast of others' reactions can itself sustain costly avoidance, despite repeated feedback and profitable opportunities to deviate.

The consequences also depend on who shares the resource. Machines can improve human coordination \citep{shirado2017} and cooperate with people at levels rivaling human cooperation \citep{crandall2018}. However, people readily exploit machines they expect to cooperate \citep{karpus2021}. Human--AI combinations also often perform worse than the better of humans or AI alone \citep{vaccaro2024}. If people adjust to aligned agents, collective costs may fall while the remaining burden is concentrated on one participant type. Measuring group averages alone would miss this redistribution.

Here we manipulate a daily broadcast while holding the game, model architecture and stated objective fixed. Adding a warning made the primary model population crowd one road despite profitable opportunities to switch. Broadcast changes and selective exposure test dependence on the information environment; other model families, a newer GPT model and reasoning settings test model dependence. Experiments with 480 participants in 36 groups show a different human response: all-human groups remain near balance, while humans sharing roads with agents increasingly use the road those agents avoid. Lower collective costs coexist with unequal costs for human and agent seats.

\section*{Results}

\subsection*{A congestion game with four levels of shared information}
We consider a basic model of resource sharing among decision makers who receive the same public information. If $n$ of $\Nagents$ commuters use a road, its travel time is $20 + 80\,n/\Nagents$ minutes: 60 min under equal use, approaching 100 min when almost everyone chooses it. The computational core used 50 agents, each implemented as a separate call to the primary model snapshot (\texttt{gpt-5.4-mini-2026-03-17}; Materials and Methods). In round 1, routes were assigned randomly; agents began choosing in round 2. Each prompt contained the game rules, a mild individual time-sensitivity trait, the agent's own road and travel time on the previous five days, and the current broadcast. Every condition instructed agents to consider how others might react to the same information. F0 supplied no report; F1 reported both roads' travel times from the previous day; F2 named the previously less-crowded road; and F3 added a warning to F2: ``However, many drivers are expected to see this same information and switch to Route B, so Route B may become congested'' (with the route label updated by the environment). The human experiments used the F1 and F3 broadcasts verbatim.

For a single run or room, $\hat p_t$ is the realized share on physical Route A in round $t$, and $\hat f_t$ is the realized share choosing the previously less-crowded road. Hats distinguish observed fractions from choice probabilities. We write $\bar f=\mathbb{E}_t[\hat f_t]$ for the within-run time average over eligible rounds, and $\fmean$ for the equal-weight mean of these averages across runs or rooms. The primary summaries use rounds 6--60: route imbalance $I$, the mean absolute deviation of a road's population share from one half; switching fraction $M$, the mean fraction changing roads between rounds; and realized mean travel time. We report medians of run-level $I$, $M$ and travel time unless stated otherwise; $\fmean$ always denotes a mean across runs. Imbalance ranges from 0 under equal use to $\tfrac12$ when everyone uses one road. Mean travel time is $\mathbb{E}_t[T_t]=60+160\,\mathbb{E}_t[(\hat p_t-\tfrac12)^2]$, where $\mathbb{E}_t$ averages rounds within one run, not across runs (Materials and Methods). This cost penalizes both persistent concentration and repeated movement between crowded roads.

\begin{figure}[tbp]
\centering
\includegraphics[width=\textwidth,height=0.72\textheight,keepaspectratio]{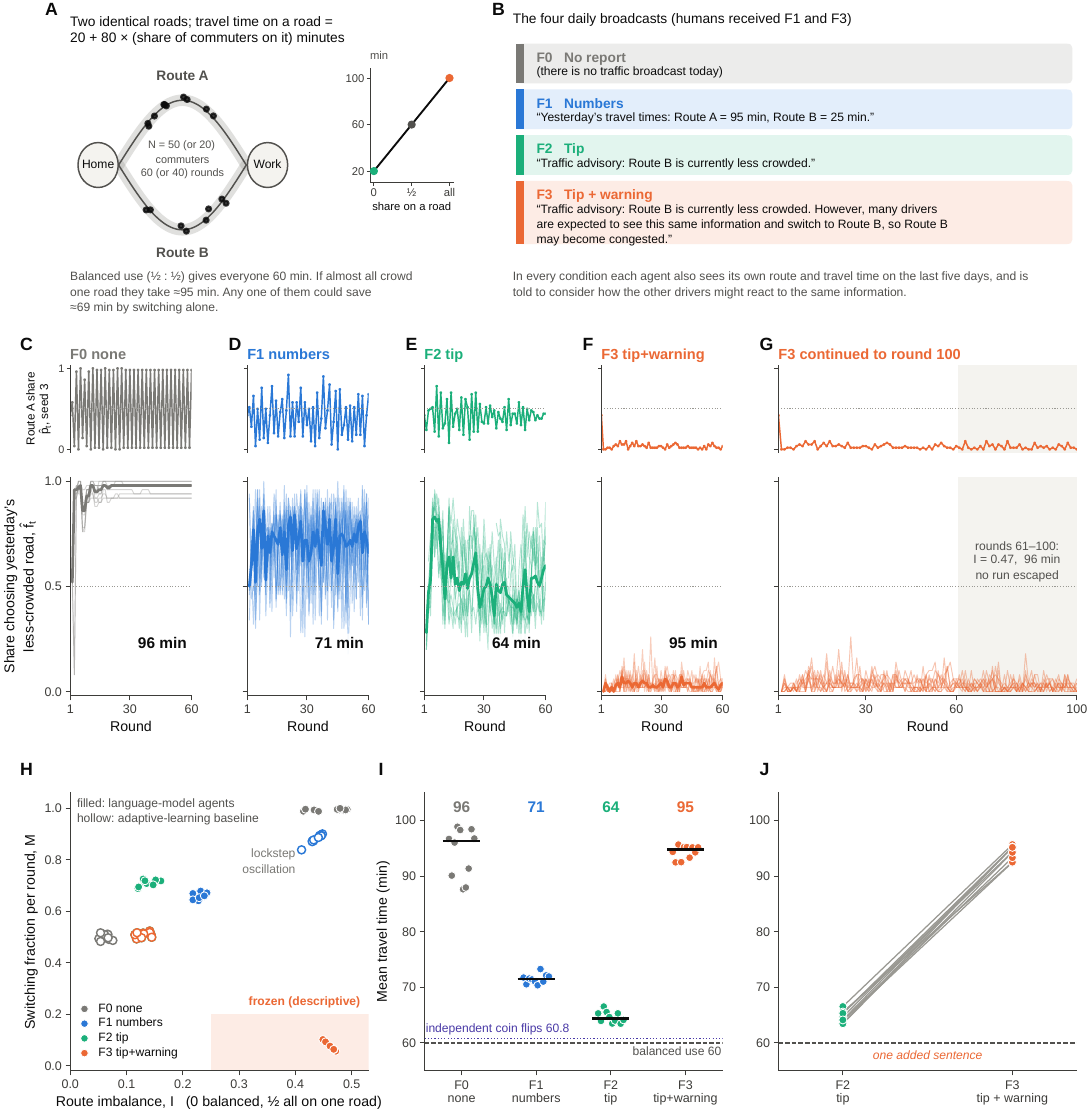}
\caption{\textbf{A shared warning produces persistent concentration in the primary model population.}
(\textbf{A}) The game: two identical roads, $\Nagents$ commuters choosing simultaneously every round, travel time increasing linearly with a road's share of commuters (inset). (\textbf{B}) The four daily broadcasts (verbatim). (\textbf{C} to \textbf{F}) For each condition of the prespecified core (50 GPT agents, 60 rounds, ten seeds), the upper strip shows the share of agents on physical Route A in one representative run (seed 3): lockstep oscillation under no report, damped oscillation under numbers, near-balance under the tip, and persistent concentration on one road under the tip plus warning. The main panel shows the share choosing the road that was less crowded in the previous round, $\hat f_t$, for all ten runs (thin lines; thick line, median; number, median of run-mean travel time). (\textbf{G}) The ten F3 runs continued to round 100 (shaded). (\textbf{H}) Route imbalance $I$ against switching fraction $M$ for all 40 runs (filled) and for the multinomial-logit learning baseline on the same seeds (hollow); the shaded region marks the descriptive frozen criterion. (\textbf{I}) Mean travel time per run (bars, medians; dashed, balanced use; dotted, expectation for independent coin flips). (\textbf{J}) Paired F2$\rightarrow$F3 change in mean travel time by seed.}
\end{figure}

\subsection*{One warning produces persistent concentration}
The prespecified computational core compared four conditions across ten paired seeds (40 runs of 50 agents and 60 rounds; 118{,}000 model decisions). Without a broadcast, agents still received their own travel-time history. In the representative F0 run (Fig.~1C), the random 21:29 split became 29:21 in round 2. All 29 agents on the slower road then switched, while 19 of the other 21 stayed, producing a 2:48 split. Subsequent near-unanimous switching moved the crowd between roads rather than balancing it. Thus, feedback through experienced congestion can align choices even without a public report. Across runs, almost every agent changed roads each morning ($M = 0.99$); populations alternated between highly unequal splits, giving mean travel time of 96 min (Fig.~1C, H, I). Numerical reports reduced imbalance and switching ($I = 0.23$, $M = 0.66$, 71 min; Fig.~1D). The bare tip gave the lowest cost among these four conditions ($I = 0.13$, 64 min; Fig.~1E). The references are 60 min under equal use and 60.8 min under independent fair choices (Fig.~1I): random round-to-round imbalances put more commuters on the crowded road, raising the expected cost above 60 min. Adding the warning produced persistent concentration: about 96\% of agents used the same road, with $I = 0.47$, $M = 0.07$ and 95 min (Fig.~1F). Most agents kept choosing the previously crowded road, consistent with anticipating that others would move to the advertised alternative. Their stated reasons frequently invoked others switching or crowding (Fig.~2H), although these reports do not independently establish internal reasoning. The paired F3--F2 contrast was $+0.32$ in $I$ (minimum over seeds $+0.30$) and $-0.63$ in $M$ (maximum $-0.61$), with all ten pairs in the prespecified directions (claim-level sign tests with Holm adjustment, $p = 0.0039$); mean travel time rose by 30 min (range 27--32 min; Fig.~1J).

For a learning-based comparison, we also specified agents that choose by multinomial logit: they probabilistically favor roads with lower travel-time estimates, updated from experience and available numerical reports (Materials and Methods). Adding numerical reports (F0$\rightarrow$F1) had opposite effects in the two populations. It reduced imbalance and switching in GPT populations ($I$: 0.48 to 0.23; $M$: 0.99 to 0.66), but increased both in the learning baseline ($I$: 0.06 to 0.44; $M$: 0.50 to 0.88), producing oscillation (Fig.~1H; SI Appendix, Table~S1). The prespecified interaction tests this difference between the two F0-to-F1 changes; its direction held in all ten paired seeds for both outcomes (SI Appendix, section S4). The baseline treats the bare tip (F2) and tip plus warning (F3) identically because it has no representation of the warning sentence.

Persistent concentration left large unilateral gains available. At a 47:3 split, an agent on the crowded road takes 95.2 min and could reduce its travel time to 26.4 min by switching alone, holding everyone else's choice fixed. Agents rarely took this opportunity ($M=0.07$); the concentrated state is therefore not a Nash equilibrium. None of the ten F3 populations escaped when continued to round 100: rounds 61--100 had $I = 0.47$ and 96 min (Fig.~1G). We call a run ``frozen'' when it combines high imbalance, low switching and a persistent majority road under the prespecified descriptive classifier (Materials and Methods). All ten F3 runs and none of the 30 other core runs met this criterion.

\begin{figure}[tbp]
\centering
\includegraphics[width=\textwidth,height=0.72\textheight,keepaspectratio]{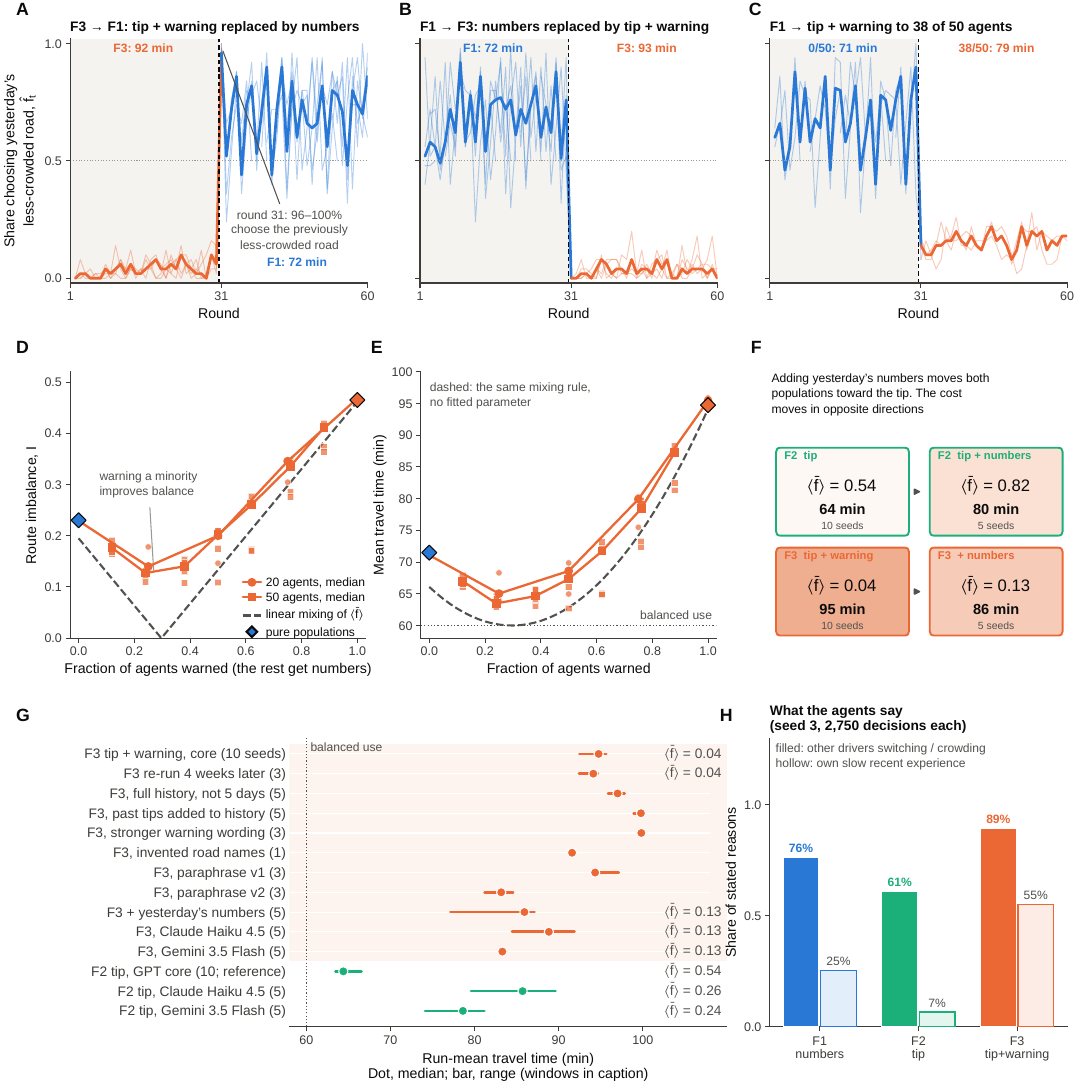}
\caption{\textbf{The collective state follows the message currently in force.}
(\textbf{A} to \textbf{C}) $\hat f_t$ when the broadcast is changed at round 31 (dashed line): the tip plus warning replaced by numerical reports (F3$\rightarrow$F1; A, five seeds), numerical reports replaced by the tip plus warning (F1$\rightarrow$F3; B, five seeds), and the warning given to 38 of 50 agents while the rest keep receiving numbers (C, three seeds); numbers give mean travel time in rounds 16--30 and 36--50. In A, 96--100\% choose the previously less-crowded road in the first post-switch round; 86--98\% actually change roads, and congestion remains high before costs subsequently decline. (\textbf{D}, \textbf{E}) Imbalance and mean travel time against the fraction of agents receiving the warning while the rest receive numbers (points, runs; lines, medians; diamonds, pure populations); dashed curves, linear mixing of the two pure populations' tip-following shares with no fitted parameter, which approximates the shape but under-predicts cost near the minimum. (\textbf{F}) Appending yesterday's travel times to the tip or to the tip plus warning: $\fmean$ and the median run-mean travel time per cell. (\textbf{G}) Run-mean travel time (dot, median across runs; bar, range) for message, history and model variants; windows are rounds 6--40 for the 20-seat history variants and 6--60 for the remaining runs. $\fmean$ is shown where round-level data are available. (\textbf{H}) Share of agents' stated reasons that invoke other drivers switching to or crowding a road (solid) or the agent's own slow recent commute (light), seed 3, 2{,}750 decisions per condition (coding rules in Materials and Methods).}
\end{figure}

\subsection*{Changing the broadcast changes the collective dynamics}
We changed the broadcast at round 31 in fresh runs to test whether the established pattern could be disrupted (Fig.~2A--C). Replacing the tip plus warning with numerical reports (F3$\rightarrow$F1) immediately increased switching in all five seeds. In round 31, 96--100\% of agents chose the previously less-crowded road, while 86--98\% changed their own road. Concentration consequently persisted on the opposite road: $I = 0.46$--$0.50$ and mean travel time was 94--100 min. Costs subsequently declined. In the prespecified post-intervention window (rounds 36--50), mean travel time was 67--74 min, compared with a pre-intervention mean of 92 min; stationary F1 runs gave 70--74 min over the same window. Replacing numbers with the tip plus warning (F1$\rightarrow$F3) induced avoidance immediately: no agent chose the previously less-crowded road in round 31, and rounds 36--50 gave 91--96 min, compared with 72 min before the change. Switching 38 of 50 agents from F1 to F3 while retaining F1 for the other 12 raised travel time from 71 to 79 min. Thus, the current broadcast could override an established collective state: concentration was neither irreversible nor determined solely by the initial allocation. These switches do not establish weak initial-condition dependence in general. These interventions changed both the warning and the information format; they do not isolate removal of the warning sentence alone.

\subsection*{Dose--response, and the benefit of warning a minority}
Selective exposure to the warning produced a non-monotonic response (Fig.~2D, E). In 20-agent populations, median $I$ at 0, 5, 10, 15 and 20 warned agents was 0.23, 0.14, 0.20, 0.35 and 0.47; corresponding travel times were 71, 65, 69, 80 and 95 min. The 50-agent grid showed a similar shape, with its lowest observed cost at 12 warned agents. Both exposure grids ran for 60 rounds. Why can warning a minority help? With numerical reports (F1), the previously less-crowded road attracts about 70\% of agents, only moderately above the balanced share of 50\%. With the warning (F3), only about 4\% choose it: the avoidance response is much stronger than the original tendency to follow. Replacing roughly 30\% of the numerical-report recipients with warning recipients therefore brings the mean share near 50\%, if each group retains its original response (SI Appendix, section S5). A linear mixture of the two pure populations' average responses approximates the U-shape and a minimum near 30\% exposure, but under-predicts imbalance near the minimum (5 of 20 warned: predicted 0.03, observed 0.14; 12 of 50: 0.04 versus 0.13). A homogeneous independent-choice benchmark at the same mean also under-predicts imbalance (0.09 and 0.06, respectively; SI Appendix, Table~S7). The mixture is therefore a descriptive approximation; it does not capture the full variability or the changes in behavior caused by mixing the two groups.

\subsection*{Numerical feedback and message variants define the effect's limits}
We next asked whether explicit numerical evidence changed the response to the warning. Yesterday's travel times were appended to the tip and to the tip plus warning (five seeds each). Three concurrent repeats of the original warning condition (F3) showed stable outcomes relative to the core (SI Appendix, Fig.~S3; Materials and Methods). Numbers increased tip-following under both messages, with different effects on cost (Fig.~2F). Added to the bare tip, they raised travel time from 64 to 80 min, and all five runs were classified as oscillating. Added to the warning, they reduced travel time from 95 to 86 min; two of five runs still met the frozen criterion. The numeric warning condition retained strong avoidance: the average tip-following share was 0.13 despite explicit feedback showing the other road's lower previous cost.

Most additional variants retained the frozen state, including full personal history, past advisories added to the history, stronger wording and invented road names (Fig.~2G). Of the two alternative phrasings of the warning, the first retained the frozen state; the second produced a partial state (83 min, $M = 0.25$). Claude Haiku and Gemini also failed to meet the complete frozen criterion, as did the primary model with reasoning enabled (detailed below). These differences identify boundaries of the effect rather than uniform robustness. In a descriptive analysis of 2{,}750 stated reasons from core F3 seed 3, 89\% referred to other drivers switching or crowding, 55\% referred to the agent's own slow experience, and 50\% contained both codes (Fig.~2H, Fig.~3F; SI Appendix, section S4). These overlapping categories describe generated explanations, not the model's internal computation.

\begin{figure}[tbp]
\centering
\includegraphics[width=\textwidth,height=0.72\textheight,keepaspectratio]{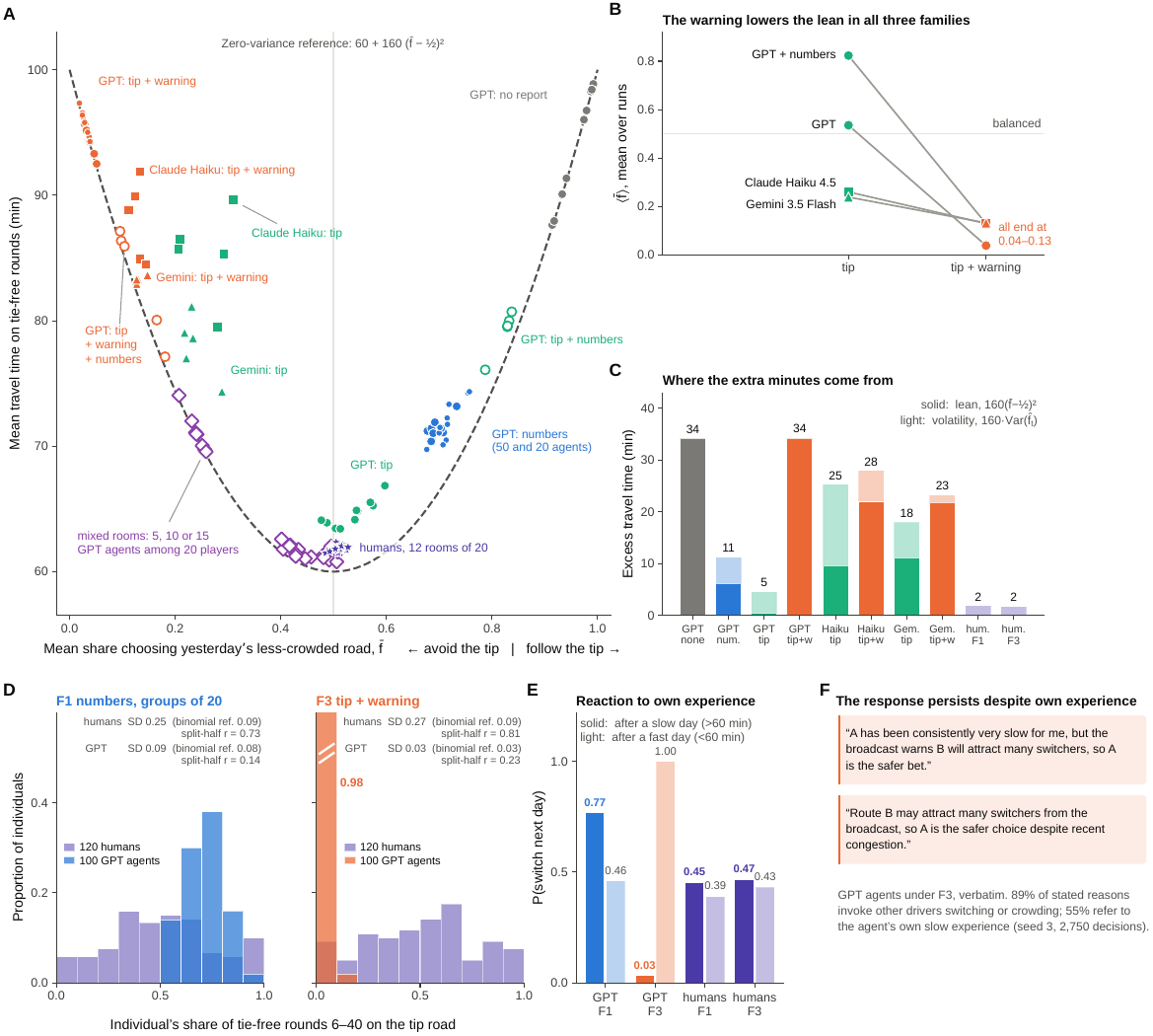}
\caption{\textbf{Average response and temporal variation contribute to cost, while individual response patterns differ.}
(\textbf{A}) Mean travel time on tie-free analysis rounds against $\fbar$ for the core, matched and cross-provider/numeric all-agent runs (color, message; marker, model family), for the twelve all-human rooms (stars) and for the 24 mixed rooms of Fig.~5 (hollow diamonds; room-level share over all seats). The dashed curve is the zero-variance reference from Eq.~\ref{eq:cost}. Both axes use the same rounds, so each point lies above it by exactly $160\,\mathrm{Var}_{\mathcal R}(\hat f_t)$. (\textbf{B}) $\fmean$ under the tip and under the tip plus warning for each model family (five to ten runs per point): the warning lowered average tip-following in all three families, from different baselines. (\textbf{C}) Decomposition of excess travel time into the average-response component, $160(\fbar-\tfrac12)^2$, and temporal variation, $160\,\mathrm{Var}(\hat f_t)$; components are calculated within each run or room and then averaged with equal weight. For Claude and Gemini, their opposing changes reduce the net cost increase. (\textbf{D}) Distribution of individuals' tendency to follow the tip (share of tie-free rounds 6--40), as a proportion of individuals, for 120 humans (20 $\times$ six rooms) and 100 GPT agents (20 $\times$ five runs) per condition in 20-seat groups; annotations give the SD, the SD expected for identical choosers at the same mean, and the split-half correlation of individual tendencies. (\textbf{E}) Mean individual probability of switching after a slow ($>60$ min) or fast ($<60$ min) day, for transitions from rounds 6--39 to 7--40. Individuals with no qualifying previous outcome are omitted from that conditional mean. (\textbf{F}) Verbatim reasons given by GPT agents under F3.}
\end{figure}

\subsection*{Average response and temporal variation account for cost differences}
Let $\hat f_t$ be the fraction choosing the road that was less crowded in round $t-1$. Under the bare tip (F2) and tip plus warning (F3), this is the road explicitly named in the broadcast; under numerical reports (F1), it is the road with the lower reported travel time. To measure the response to a previous imbalance, we use rounds whose preceding split was unequal, denoted $\mathcal R$. A preceding tie leaves no uniquely less-crowded road, so that observation is excluded from $\hat f_t$ summaries; a balanced outcome in the current round is retained. No such preceding ties occurred in the core warning runs; they were infrequent in the computational runs but more common in the nearly balanced human rooms (SI Appendix, sections S4 and S5). Since $\hat p_t$ equals either $\hat f_t$ or $1-\hat f_t$, applying the identity $\mathbb{E}[(\hat f_t-\tfrac12)^2]=(\mathbb{E}[\hat f_t]-\tfrac12)^2+\mathrm{Var}(\hat f_t)$ to the round cost gives
\begin{equation}
\mathbb{E}_{\mathcal R}[T_t] = 60 + 160\,\big[(\mathbb{E}_{\mathcal R}[\hat f_t]-\tfrac12)^2 + \mathrm{Var}_{\mathcal R}(\hat f_t)\big].
\label{eq:cost}
\end{equation}
Here $\mathbb{E}_{\mathcal R}$ averages equally over the retained rounds within one run or room; $\fbar=\mathbb{E}_{\mathcal R}[\hat f_t]$. The two terms separate a sustained bias in the average response from variation in that response across rounds. Figure~3A plots travel time and $\fbar$ on the same tie-free rounds; the zero-variance curve is a lower reference, and each run's vertical distance from it is exactly the variance contribution. Equation~\ref{eq:cost} follows from the cost function and applies to any decision rule. All-round outcomes remain the primary summaries; their difference from the tie-free summaries is reported in SI Appendix, section S5.

Under F0, the primary model almost always chose the previously less-crowded road ($\fmean = 0.96$). Although physical road occupancy oscillated, $\hat f_t$ was nearly constant: its excess cost was almost entirely the average-response term (34.1 of the 34.2 excess minutes; Fig.~3C). Numerical reports gave $\fmean = 0.69$, with approximately 6 min from the average-response term and 5 min from temporal variation. The bare tip gave $\fmean = 0.54$, with 0.4 min and 4 min from the two terms. The warning shifted $\fmean$ to 0.04, again with little temporal variation. Adding numbers increased $\fmean$ under both messages (0.54 to 0.82 and 0.04 to 0.13). These decompositions are calculated for each run and then averaged; they distinguish variability in $\hat f_t$ from switching between physical roads.

In an extension planned after the core results, Claude Haiku 4.5 and Gemini 3.5 Flash were each tested under F2 and F3 at five paired seeds (Fig.~3B, C; SI Appendix, Fig.~S2). The warning reduced switching in all ten pairs (mean $\Delta M = -0.15$ for Claude and $-0.16$ for Gemini), increased imbalance in three Claude pairs and all five Gemini pairs, and raised mean travel time by approximately 3 and 5 min, respectively. None met the frozen criterion. In both families the warning reduced $\fmean$ (Claude, 0.26 to 0.13; Gemini, 0.24 to 0.13). The average-response cost increased by 12.5 and 10.7 min, while the temporal-variation cost fell by 9.8 and 5.4 min. All pairs showed this opposing movement of the two cost components. Thus, the direction of the warning response extended across the three tested families, whereas the full frozen state and the size of the cost increase depended on the model.

\subsection*{Reasoning changes the collective outcome but does not consistently remove avoidance}
A second extension varied the reasoning setting and the model generation at the same five seeds (SI Appendix, section S9, Fig.~S10 and Tables~S13 and S14). Without reasoning, raising the primary model's temperature from 0.7 to 1.0 left the core contrast essentially unchanged. All five F3 runs were frozen, while no F2 run was; the warning added 27.6 min to mean travel time. With low or medium reasoning effort, the warning still increased imbalance, reduced switching and raised travel time in all ten pairs. No run met the frozen criterion, however. Under the warning, reasoning agents chose the previously less-crowded road more often ($\fmean = 0.26$ and $0.21$, against $0.06$ without reasoning). Their mean travel time was 70.0 and 74.7 min, against 91.5 min (means over five runs). GPT-6 Luna without reasoning also moved toward avoidance under the warning in all five pairs, without freezing. At its default medium reasoning, Luna froze under both messages (ten of ten runs; $\fmean = 0.05$ under the tip and 0.07 under the warning; 92.4 and 89.5 min). In this setting the bare tip was enough to produce the costly avoidance. Adding the warning made no consistent difference. The warning thus moved responses toward avoidance wherever avoidance was not already complete. Whether the population froze depended on the model and on its reasoning setting.

\begin{figure}[tbp]
\centering
\includegraphics[width=\textwidth,height=0.72\textheight,keepaspectratio]{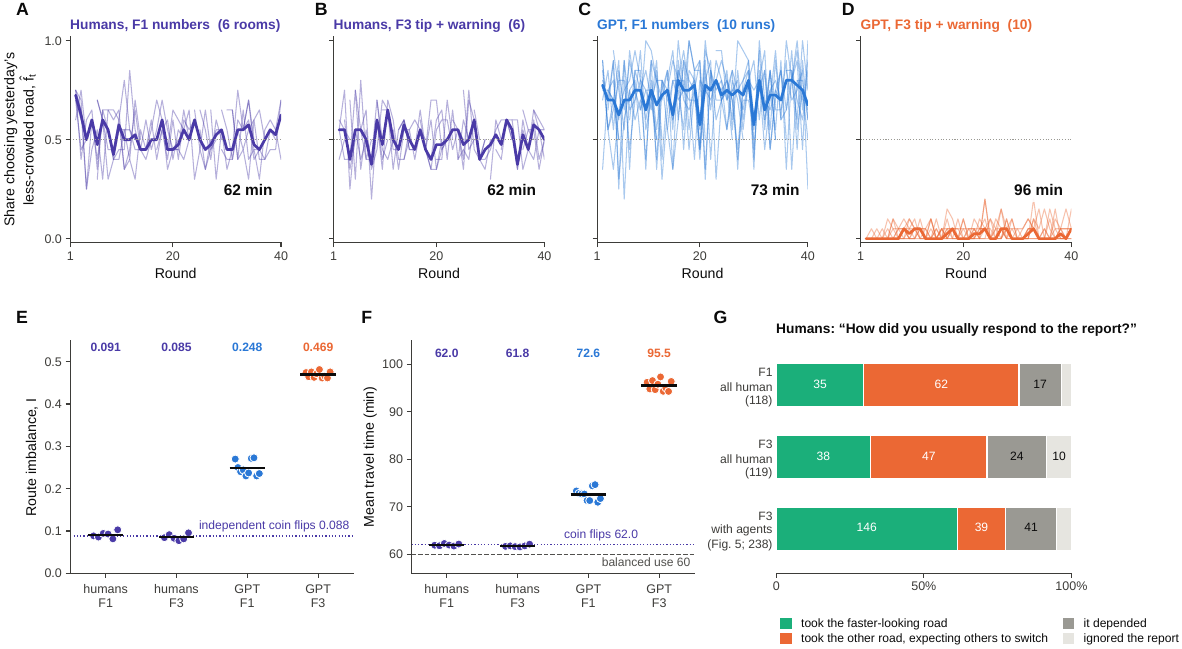}
\caption{\textbf{Humans in the same game.}
(\textbf{A} to \textbf{D}) $\hat f_t$ in six human rooms per condition (Prolific, 20 participants, 40 rounds) and in ten GPT runs of the same size; thin lines, rooms or runs; thick lines, round-wise medians; inset travel times, means across rooms or runs. Gaps mark rounds in which the previous counts were equal. (\textbf{E}, \textbf{F}) Imbalance and mean travel time per room or run (points; bars, means); dotted, expectation for independent coin flips; dashed, balanced use. (\textbf{G}) Participants' self-reported usual response to the report in the all-human rooms and in the mixed rooms of Fig.~5.}
\end{figure}

\subsection*{Human groups remain near balance under both tested broadcasts}
Two hundred forty participants recruited through the online experiment platform Prolific (\url{https://www.prolific.com}) played the 20-seat, 40-round game in twelve rooms, six under numerical reports (F1) and six under the tip and warning (F3). They received the same broadcasts and five-day personal history as the agents, with timed decisions. A fixed payment was supplemented by a bonus that increased as their own mean game travel time over 40 rounds decreased (Materials and Methods). All twelve rooms remained near balance: $I = 0.09$ under each message and mean travel time of 62.0 and 61.8 min, respectively (Fig.~4A, B, E, F). These values are close to the independent fair-choice references of 0.088 and 62.0 min for 20 seats. The registered descriptive blocked F3--F1 estimate was $-0.005$ in $I$ (two-sided $p = 0.28$; constant-shift 95\% interval $-0.016$ to $+0.006$) and $+0.03$ in $M$ ($p = 0.31$). The intervals and the small number of rooms do not establish equivalence. GPT populations matched for group size and duration incurred 73 and 96 min under the two messages (means across runs) (Fig.~4C, D). This comparison is contextual: the human and model studies were non-concurrent and differed in implementation and incentives.

Most participants reported using the broadcast: 97 of 118 respondents under numerical reports (F1) and 96 of 119 under the warning (F3) said it influenced them moderately or strongly. Asked about their usual response, 35 and 38 reported taking the faster-looking road, whereas 62 and 47 reported taking the other road in anticipation of others switching (Fig.~4G). These retrospective reports are consistent with varied response tendencies, but do not independently identify the process that maintained balance.

\subsection*{Individual response patterns differ between human and model populations}
Individual response patterns differed between the populations (Fig.~3D). We measured each person's or agent's share of tie-free rounds 6--40 spent on the previously less-crowded road, using 120 humans (20 participants $\times$ six rooms) and 100 GPT agents (20 agents $\times$ five runs) per condition. GPT tendencies were concentrated under numerical reports (F1; mean 0.71, SD 0.09) and near zero under the warning (F3): 98 of 100 agents followed the tip on fewer than 10\% of eligible rounds (mean 0.03, SD 0.03). Their dispersion was close to a homogeneous independent-choice reference at the observed mean (SD 0.08 and 0.03, respectively). Correlations between individuals' first- and second-half tendencies were modest ($r = 0.14$ and 0.23).

Human tendencies were more dispersed (SD 0.25 and 0.27, compared with 0.09 under the reference) and more reproducible across session halves ($r = 0.73$ and 0.81). Both consistent followers and avoiders were present, while mean tip-following was 0.51 under each message. Seven of 120 participants under numerical reports (F1) and 11 of 120 under the warning (F3) followed the tip on fewer than 10\% of eligible rounds. These between-condition distributions do not establish how the warning changed any particular individual's strategy. The reference and split-half analyses are descriptive: observed tendencies combine persistent individual differences, learning and shared room history.

Responses to recent outcomes also differed (Fig.~3E). For GPT agents under numerical reports, the mean individual probability of switching after a slow day ($>60$ min) was 0.77, compared with 0.46 after a fast day ($<60$ min). Under the warning these probabilities were 0.03 and 1.00: even the occasional agent reaching the less-crowded road typically returned to the crowded road on the following round. Human switching probabilities were 0.45 and 0.39 under numerical reports (F1), and 0.47 and 0.43 under the warning (F3). These summaries average individual conditional probabilities over individuals with at least one qualifying outcome; they describe responses to experienced payoffs without directly measuring beliefs.

\begin{figure}[tbp]
\centering
\includegraphics[width=\textwidth,height=0.72\textheight,keepaspectratio]{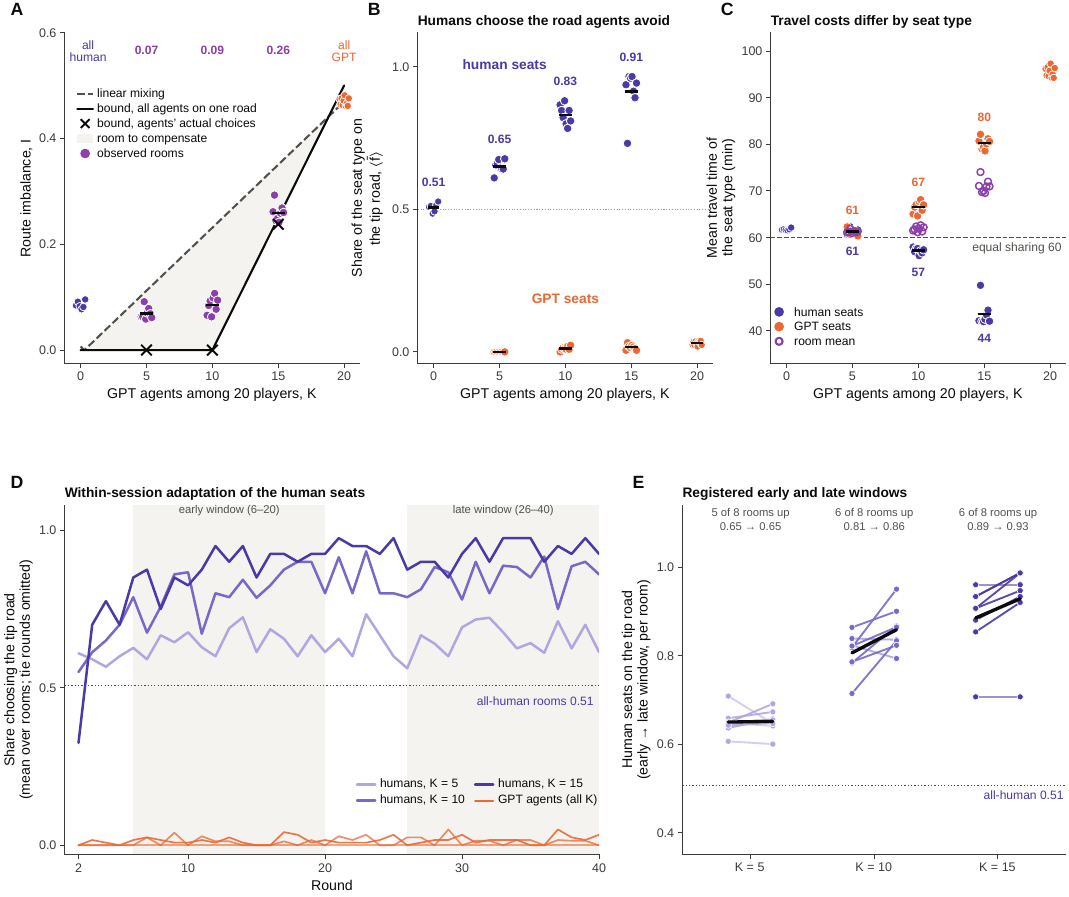}
\caption{\textbf{Human choices change within mixed rooms, with unequal costs by seat type (24 rooms).}
(\textbf{A}) Route imbalance in rooms of 20 with $K$ GPT agents, all under F3 (purple, one point per completed room; bars, means), against the prediction if humans kept their all-human lean (dashed; linear mixing), the lower bound if all agents sit on one road (solid; accounting bound), and the bound given the agents' actual choices (crosses; mean per $K$); non-concurrent all-human and all-agent reference groups are shown at $K = 0$ and $K = 20$. (\textbf{B}) Tip-following share of the human seats and of the agent seats in the same rooms (tie-free rounds 6--40). (\textbf{C}) Mean travel time of the human seats, of the agent seats and of the room; dashed, the 60 min of equal sharing. (\textbf{D}) Share of human seats (purple, by $K$) and of agent seats (orange) choosing the tipped road, round by round, averaged over rooms. (\textbf{E}) Human tip-following per room in the specified early (rounds 6--20) and late (26--40) windows; black segments join composition means. Panels summarize all eight rooms per composition; registered block-level inference and dependence sensitivities are reported in the text.}
\end{figure}

\subsection*{Human choices shift in mixed groups, with unequal costs by seat type}
We next examined groups containing $K = 5$, 10 or 15 GPT agents among 20 seats, all under F3. Two references aid interpretation. If the humans and agents retained their all-human and all-agent mean tip-following shares (0.51 and 0.03), linear mixing would give approximate imbalances of 0.11, 0.23 and 0.35 (Fig.~5A, dashed line). If all $K$ agents chose one road, the smallest achievable imbalance would instead be $\max(0,K/20-\tfrac12)$, irrespective of human choices (solid line). The latter is a conditional lower bound, equal to 0.25 at $K=15$; we also calculate it from the agents' actual round-by-round choices (crosses; SI Appendix, section S5).

The study yielded 24 valid rooms in eight blocks of three compositions (240 humans and 240 agent seats). Participants were told that some players were AI-controlled, but not their number or identities; agent prompts did not disclose human participation. The design, inclusion rules and registered analysis are summarized in Materials and Methods, with the registration and execution record in SI Appendix, section S8.

Mean room imbalance was $I=0.07$ at $K = 5$ (eight rooms, range 0.06--0.09), $I=0.09$ at $K = 10$ (eight rooms, 0.06--0.11) and $I=0.26$ at $K = 15$ (eight rooms, 0.24--0.29). These means were below the fixed-response mixture values of 0.11, 0.23 and 0.35. At $K=15$, all agents selected the same road in 80\% of analysis rounds. The lower bound conditional on their actual choices averaged 0.238, compared with 0.259 observed: human choices brought the mean imbalance within 0.022 of the best allocation permitted by those agent choices. Five of these eight rooms met the frozen classifier; all eight retained the same majority road throughout the analysis window, but three fell below its imbalance threshold.

Imbalance increased with agent share in every block, by a mean of 0.0951 per five additional agents; switching decreased in every block, by 0.2080 on average. In the registered one-sided sign-flip test over the eight block slopes, $p = 1/256 = 0.0039$ for imbalance and for switching, the smallest value attainable with eight blocks. The inference assumes independent blocks with errors symmetric about zero. Sensitivity analyses that flip blocks sharing recruitment windows jointly have only 16 and 4 sign configurations; the observed configuration was the most extreme in both ($p = 0.0625$ and $0.25$, the minimum attainable values; SI Appendix, section S8). Leaving out any one block and applying the specified nuisance sensitivities retained the directions of both slopes (SI Appendix, Tables~S10 and S11).

Agents continued to avoid the tip road in mixed rooms, with mean tip-following between 0 and 0.02 by composition (Fig.~5B). The advisory repeatedly named the same road at high agent shares: 94\% of consecutive rounds at $K=10$ and 100\% at $K=15$, compared with 54\% at $K=5$. Human tip-following averaged 0.65, 0.83 and 0.91 at $K = 5$, 10 and 15. Within sessions, it rose from 0.64 and 0.62 in rounds 2--5 to 0.87 and 0.95 in rounds 31--40 at $K=10$ and 15 (Fig.~5D). Comparing the prespecified early and late windows (rounds 6--20 and 26--40), human tip-following rose in six of eight $K=10$ rooms and six of eight $K=15$ rooms (two were unchanged); the $K=5$ mean changed little (0.65 to 0.65; Fig.~5E).

Across completed mixed rooms, 86 of 240 participants followed the tip on at least 90\% of eligible rounds, including 30 of 40 at $K=15$; the corresponding count in all-human F3 rooms was 9 of 120. Mean individual switching probabilities were 0.73 after a slow day and 0.22 after a fast day. In the questionnaire, 146 of 238 respondents reported taking the faster-looking road and 39 the other road (Fig.~4G). Together, the within-session changes and outcome-contingent switching are consistent with adaptation to a predictably less-crowded road. They do not distinguish learning from experienced payoffs from explicit understanding of the agents' policy.

At the end of the task, 230 participants estimated the percentage of AI agents in their room. Their estimates averaged 63\%, 67\% and 69\% when the actual percentages were 25\%, 50\% and 75\%, respectively (SI Appendix, Fig.~S9 and Table~S12). Behavioural adaptation therefore coexisted with substantial error in reported composition, especially when agents were a minority.

The population-level agreement with simple mean-response calculations is an accounting check. When $K\ge10$ and all agents avoid the tip, even unanimous human tip-following cannot place more than half the room on that road. Imbalance then depends linearly on the human mean response, regardless of correlations between people. Consequently, matching this reconstruction cannot establish independent human choice (SI Appendix, section S5).

Collective and individual costs differed substantially (Fig.~5C). Room-mean travel time was 61.3, 61.9 and 71.0 min at $K = 5$, 10 and 15, compared with about 96 min in the all-agent reference of the same size. The human seats averaged 61.3, 57.2 and 43.5 min and the agent seats 61.2, 66.5 and 80.2 min. Thus, at $K=15$, human seats averaged 16.5 min below the 60-min equal-sharing benchmark, while agent seats averaged about 20 min above it. Of the 240 participants, 129 averaged less than 60 min; no agent did. Lower collective costs in mixed rooms therefore coexisted with a persistent cost disadvantage for agent seats at $K=10$ and $K=15$. At $K=5$, both seat types averaged about 61 min.

\section*{Discussion}
A shared warning sustained a costly collective response in the primary language-model population. Agents repeatedly avoided the previously less-crowded road despite experiencing long travel times on the other road. The warning's addition was sufficient to change the core outcome, and explicit numerical feedback weakened, but did not eliminate, avoidance. Changing the broadcast from the tip plus warning to numerical reports disrupted the fixed-road pattern immediately; costs fell in subsequent rounds. Two other model families also shifted toward avoidance, although none of their runs met the complete frozen criterion. The primary model behaved similarly when reasoning was enabled: the warning still moved agents toward avoidance, but no population froze. A newer model at its default reasoning setting froze even under the tip alone. The common behavioral direction and the different collective outcomes make the response to shared information, rather than a universal frozen state, the main result.

We interpret the persistent avoidance as a consequence of model homogeneity interacting with shared information. A common model supplies similar decision rules, so a forecast about other drivers can elicit the same anticipatory response across many recipients, even when their personal histories differ. The warning then discourages the very migration it predicts, while repeated broadcasts sustain the response. The concentration of individual avoidance tendencies and the changes under selective exposure are consistent with this account. Homogeneity need not produce concentration under every message: populations of the same model stayed near balance under the tip alone (64 min). The proposed mechanism is the alignment induced by a shared warning across a population with similar decision rules. In GPT-6 Luna at its default reasoning setting, the tip alone was enough to produce this alignment. Its stated reasons referred to other drivers switching or crowding in 88\% of decisions at seed 3, against 61\% for the primary model under the tip.

The cost decomposition helps explain the differences between model families. Bias in the average tip-following share and variation in that share across rounds both increase travel time. For Claude and Gemini, the warning increased the first contribution while reducing the second, limiting the net cost increase. Warning only a minority also reduced costs relative to uniform numerical reports or uniform warnings. These results suggest that evaluation should consider the distribution of responses to shared information: the same message can have different effects depending on the population's starting behavior and on who else receives it.

Human groups showed a different pattern. Individual tendencies were dispersed and reproducible across session halves, while the group means remained close to balance under numerical reports and warning-bearing advisories. These dispersed tendencies offer one plausible route to balanced use; the present design does not isolate their contribution. Across the eight registered blocks, imbalance rose and mobility fell with agent share; the registered inference assumes independent symmetric block errors. In mixed rooms, we also observed a change within sessions. Human choices increasingly favored the road avoided by the agents, accompanied by a tendency to retain fast choices and leave slow ones. This is consistent with learning from payoffs without requiring explicit knowledge of the agents' policy. At $K=15$, the observed allocation approached the lower bound permitted by the agents' choices, but human and agent seats still averaged 44 and 80 min, respectively. The room average concealed this difference.

The experiments extend work on algorithmic monoculture and routing behavior \citep{kleinberg2021,ballestero2026,goodyear2025,wangisci2025} by manipulating a shared anticipatory message in an environment with endogenous costs and examining adaptation in mixed human--agent groups. The collective response arises without direct communication, although choices remain coupled through congestion. This differs from conventions developed through explicit agent interactions \citep{ashery2025}.

These findings suggest three priorities for evaluating agents that share resources. First, test populations: plausible individual responses need not produce efficient collective outcomes. Second, manipulate message content and distribution, which changed performance here without changing the underlying model. Third, report costs by participant type alongside group averages. Mixed groups had lower collective costs than the all-agent reference, yet agent seats bore higher costs than human seats at high agent shares. If clients bear the costs incurred by their agents, better aggregate performance could therefore conceal a disadvantage for those who delegate. This experiment measures costs assigned to seats in a game; testing that deployment hypothesis requires comparisons under alternative messages and exposure patterns.

Several limits define the scope of the findings. The game has two symmetric roads and immediate feedback. All prompts asked agents to consider how other drivers might react, which may favor anticipatory responses. The complete frozen state appeared in the primary model without reasoning and in GPT-6 Luna at its default reasoning setting. It did not appear in the two other families or in the primary model with reasoning enabled. Populations combining different models under the same message would test whether model diversity reduces the observed alignment. The core used no reasoning; the later five-seed extension varied request settings without inferring internal reasoning from generated explanations or token counts. The switch intervention changed numerical content as well as the warning. Continuing the same established state under F3, F2 and F1 would isolate those contributions more directly, while message controls could distinguish the forecast's content from trust in its source. Human and all-agent benchmarks were non-concurrent and differently incentivized. The mixed-group sample is complete, but composition was associated with recruitment order and some rooms were completed under amendments made after initial outcomes were known. Shared recruitment windows and nights limit the independence of the block comparisons (SI Appendix, section S8). Independent replication across days with random allocation of ready participants would strengthen causal interpretation of composition effects.

Within these limits, the study shows how shared predictions can sustain inefficient behavior and how adaptation by other decision makers can change who bears its costs. Assessing shared resources therefore requires both a population view of efficiency and a participant view of outcomes.

\section*{Materials and Methods}
\subsection*{Congestion game}
Two identical roads serve $\Nagents$ commuters, with travel time $t(n)=20+80n/\Nagents$ for a road carrying $n$ commuters. Weighting each road's travel time by the fraction using it gives the round mean:
\[
T_t=\hat p_t(20+80\hat p_t)+(1-\hat p_t)\{20+80(1-\hat p_t)\}=60+160(\hat p_t-\tfrac12)^2.
\]
Here $\hat p_t$ is physical Route A's share. Round 1 assigns routes randomly; decisions begin in round 2. The core, switching, exposure, cross-provider/numeric and reasoning studies use 60-round runs and, unless a pre/post window is specified, summaries over rounds 6--60. Matched 20-seat runs, human and mixed rooms, and the 20-seat history variants use 40 rounds with summaries over rounds 6--40. The persistence extension is summarized separately over rounds 61--100.

Route imbalance is $I=\mathbb{E}_t[|\hat p_t-\tfrac12|]$. Switching fraction $M$ averages the proportion changing roads, including the transition into the first analysis round. Travel time is averaged over all rounds in the specified window. For $\hat f_t$, the reference road is the one with fewer commuters in round $t-1$. When those counts are equal, $\hat f_t$ is undefined and the round is excluded from summaries derived from $\hat f_t$, but retained in $I$, $M$ and travel time. Write $\mathcal R$ for the remaining rounds. Both sides of Eq.~\ref{eq:cost} and Fig.~3A use this same set. Tie frequencies and the small differences between all-round and tie-free costs are reported in SI Appendix, sections S4, S5 and S9.

The frozen classifier, fixed before the core, requires $I\ge0.25$, $M\le0.20$ and at least ten consecutive rounds with the same majority road. It is a descriptive classification of a run, distinct from equilibrium or from any individual's tendency to switch. The SI uses $p_t$ and $f_t$ as shorthand for the same within-run realized fractions denoted here by $\hat p_t$ and $\hat f_t$.

\subsection*{Language-model agents}
Each agent seat receives a separate stateless model call each round. Client-side request logs for the core and matched primary-model runs record \texttt{gpt-5.4-mini-2026-03-17}, temperature 0.7, a maximum of 400 completion tokens, \texttt{reasoning\_effort = none} and JSON-object output. The cross-provider extension used \texttt{claude-haiku-4-5-20251001} with temperature 0.7 and \texttt{max\_tokens = 400}; its backend supplied no explicit thinking setting. Gemini calls used \texttt{gemini-3.5-flash}, temperature 0.7, \texttt{max\_output\_tokens = 400}, JSON output and a zero thinking budget, which the backend recorded as accepted in all extension calls. The reasoning and current-model extension used \texttt{gpt-5.4-mini-2026-03-17} without reasoning at temperature 1.0 and with \texttt{reasoning\_effort} low or medium, and \texttt{gpt-6-luna} with \texttt{reasoning\_effort} none or medium, its default. The reasoning-enabled requests and all Luna requests omitted temperature; no numerical value of an internal sampling temperature was independently verified. The completion cap was 400 tokens without reasoning and 4{,}000 tokens with reasoning. These are recorded request configurations, not independent measurements of internal computation. Further implementation details and the scope of the request audit are reported in SI Appendix, section S1.

The prompt (SI Appendix, section S1) gives the rules and cost schedule in words, an individual time-sensitivity trait drawn from $U(0.8,1.2)$ and rendered as one of three phrases, the agent's own road and travel time for the last five days, and the broadcast. It asks the agent to consider expected travel time and other drivers' responses, and to return a route and a one-sentence reason as JSON. A run's seed controls initial routes, traits, label mapping and baseline random draws. No provider sampling seed was supplied, so repeating a seed reproduces the specified setup rather than the exact model output. Physical roads are mapped to displayed labels by seed parity. Broadcasts are generated from previous counts; F1 reports integer travel times, F2 names the previously less-crowded road, and F3 appends the warning. Numeric variants append F1 to F2 or F3. Agents do not receive others' choices or explanations and do not communicate directly. If the previous counts tied, F1 showed 60 min for both roads; F2/F3 still named physical road 0 under the fixed tie-break, for humans and agents alike (SI Appendix, section S2).

Reason coding uses two overlapping regular-expression categories: references to others switching or crowding, and references to one's own slow experience. Percentages use all eligible reasons as the denominator; the intersection is reported separately (SI Appendix, section S4). The individual GPT analyses use seeds 3--7 of the matched F1 and F3 runs, providing 100 seats per condition. Conditional switching averages individual probabilities for transitions from rounds 6--39 to 7--40, excluding individuals without a qualifying previous outcome. This differs from $M$, whose window includes the round-5-to-6 transition.
\subsection*{Multinomial-logit learning baseline}
The comparison agents maintain estimates $v_{ir}$ of travel time on each road $r$, initialized at 60 min. After a choice, the experienced road's estimate is updated toward its observed cost with learning rate 0.3; under numerical reports (F1), both estimates are then updated toward the reported times with rate 0.5. Choices follow $P_i(r)=\exp(-\beta_i v^{\ast}_{ir})/\sum_s\exp(-\beta_i v^{\ast}_{is})$, where $\beta_i=0.10u_i$, $u_i\sim U(0.8,1.2)$, and the bare and tip plus warnings (F2/F3) both subtract 4 min from the indicated road's effective estimate $v^{\ast}_{ir}$. Other estimates are unchanged for choice. The fixed parameters and seeded starting conditions were shared across the core comparisons. This baseline represents adaptive probabilistic choice; it cannot interpret the additional warning sentence (SI Appendix, section S3).
\subsection*{Computational experiments}
\begin{itemize}[leftmargin=*,label=$\bullet$,itemsep=0.5em]
\item \emph{Core (prespecified).} Four conditions $\times$ ten seeds (3--12), 50 agents, 60 rounds; 118{,}000 decisions. Two claims were prespecified in an execution-time code freeze: L1, that $(F0-F1)_{\mathrm{LLM}}-(F0-F1)_{\mathrm{MNL}}>0$ for both $I$ and $M$, comparing the language-model population with a multinomial-logit learning baseline on the same seeds; L2, that F3 raises $I$ and lowers $M$ relative to F2. Each claim was tested by exact two-sided sign tests on the ten paired seeds, combined within a claim by the maximum $p$ and Holm-adjusted across the two claims.
\item \emph{Persistence.} The ten F3 runs were continued from their round-60 state to round 100 under F3; no run escaped the concentrated state.
\item \emph{Switching.} Five seeds each of F3$\rightarrow$F1 and F1$\rightarrow$F3 with the broadcast changed from round 31, and three seeds in which the warning was given to 38 of 50 agents from round 31; prespecified comparison window rounds 36--50 against rounds 16--30 and against stationary runs. Round-by-round values are in SI Appendix, section S6 and Table~S9.
\item \emph{Exposure.} Twenty-agent populations with 0, 5, 10, 15 or 20 warned agents (three seeds each) and 50-agent populations with 6--44 warned agents (five seeds each), the remaining agents receiving F1.
\item \emph{Robustness.} We varied history length, history of past tips, warning strength, road labels and message wording, and tested Claude Haiku and Gemini. Conditions and seed counts are in SI Appendix, Table~S4. Except for a pre-core road-name screening run, variants were specified after the core; this was not a factorial design.
\item \emph{Cross-provider and numeric extension.} Planned after the core and run about four weeks later: paired F2/F3 runs for Claude Haiku and Gemini, primary-model runs with appended numbers, and F3 repeats as a drift control (33 runs, 97{,}350 decisions). No prespecified drift flag was triggered (SI Appendix, Table~S3 and Fig.~S3). Sampling variation and changes over time cannot be separated; this extension was descriptive.
\item \emph{Reasoning and current-model extension.} Planned with earlier results known and run on 23 September 2026: paired F2/F3 runs at seeds 3--7 for the five request settings above (50 runs, 147{,}500 decisions), referenced to core seeds 3--7. A prospective plan fixed the design and descriptive analysis; no significance criterion was added (SI Appendix, section S9).
\end{itemize}

\subsection*{Human benchmark}
Both human studies used nonprobability, opt-in samples recruited through Prolific and were conducted online in a browser; no population representativeness is claimed. Repeat participation was prohibited; identifier checks confirmed 480 distinct participants across the two analyzed samples. Age and gender summaries are in SI Appendix, section S7. Two hundred forty adults were recruited through the online experiment platform Prolific (\url{https://www.prolific.com}; English-fluent, desktop, approval rate $\ge 95\%$) into twelve 20-person rooms in three blocks, two rooms under F1 and two under F3 per block by a committed within-block randomization, and played 40 rounds with timed choices and a five-round personal history (SI Appendix, section S7). Timeouts carried the previous road forward; three consecutive timeouts caused dropout with continued carry-forward, and the main summaries retain this handling. Ordinary dropout did not invalidate a room. A registered per-protocol sensitivity excludes rooms with at least two permanent dropouts first flagged by round 30; one F3 room met this criterion and is excluded only from that descriptive sensitivity (SI Appendix, section S7). Participants received a base payment of GBP 4.50 plus a bonus of GBP $2\,(100 - \mathbb{E}_t[t_i(t)])/80$ (at most GBP 2), where $\mathbb{E}_t[t_i(t)]$ is their mean travel time over all 40 rounds, and answered a closed-item questionnaire on the report's influence, their usual response to it and their understanding of the rules. The Ethics Review Committee of the Research Center for Advanced Science and Technology, The University of Tokyo, approved the study (approval E26ALS0545). All participants gave informed consent. Individual tendencies in Fig.~3D are each participant's share of tie-free rounds 6--40 on the tip road; split-half reliability correlates the tendency over rounds 6--22 with that over rounds 23--40 across individuals. The descriptive message contrast averages within-block F3--F1 differences over three blocks; exact inference is specified in SI Appendix, section S7. The benchmark preceded the mixed study and supplies only a contextual $K = 0$ reference.

\subsection*{Mixed-group experiment}
\emph{Design and analysis.} Eight blocks each contained one 20-seat room with 5, 10 and 15 GPT agents under the tip and warning (F3): 24 rooms, 240 humans and 240 agent seats. Participants knew that some players were AI-controlled, but not their number or identities. Agents used the 20-driver core prompt and were not told about human participation. Timing, history, timeout and payment rules matched the human benchmark. The same ethics approval (E26ALS0545) covered this experiment. All participants gave informed consent. The primary statistic was the mean of the eight within-block imbalance slopes, $b_b=(I_{b,15}-I_{b,5})/2$; switching fraction $M$ was supporting and seat-type decompositions were descriptive. The registered sign test, assumptions and sensitivities to dependence between recruitment windows are specified in SI Appendix, section S8. Composition was associated with arrival order because larger human quotas were filled first; this limits causal interpretation.

\emph{Execution and inclusion.} Collection on 14--17 September 2026 yielded all 24 valid rooms, each completing 40 rounds. Technical failures excluded two started attempts, whose records were preserved separately; authorized replacement attempts supplied new complete rooms. The 240 included participants produced 93 timeouts among 9{,}360 decisions in rounds 2--40 and six permanent dropouts. When a participant timed out, the system reused their previous road; after three consecutive timeouts, it continued to use that road for the remaining rounds. These system-assigned choices and their travel times were included in the primary analysis. No mixed room had two permanent dropouts by round 30. Outcomes use rounds 6--40, and early and late windows are rounds 6--20 and 26--40. Registration amendments, knowledge of preceding outcomes, recruitment and inclusion records are documented in SI Appendix, section S8 and Figs.~S7 and S8.

\emph{Composition estimates.} An optional end-of-task question asked participants to estimate the automated share of their 20-person group. This registered secondary measure was summarized descriptively; missing answers were not imputed and beliefs did not affect inclusion or primary analyses. Response recovery, weighting and missingness are detailed in SI Appendix, section S8.

\subsection*{Registration and study periods}
Human-study plans and amendments were registered on OSF (\url{https://osf.io/whsyu}). The all-human benchmark was collected on 1--6 September 2026 and the mixed-group data on 14--17 September 2026. The computational core hypotheses and classifier were fixed in an execution-time code freeze before the August runs; these computational plans were not public preregistrations. Subsequent computational extensions were planned with earlier results known. Some mixed-group completion and replacement attempts were authorized after initial outcomes were known; SI Appendix, section S8 provides the full registration and execution history. Individual-response, split-half, reason-coding and additional accounting analyses are exploratory; the registered primary definitions of $I$ and $M$ were retained.

\subsection*{Statistics}
Runs and rooms are the units for treatment comparisons. Agent-rounds and individual participants are not treated as independent treatment replicates. Confirmatory analyses comprise the two prespecified computational-core claims and the registered mixed-group composition test, the latter conditional on its stated block-error model and amended collection history. Other comparisons are descriptive. Unless specified otherwise, computational condition summaries are medians over runs, human-benchmark outcome summaries are means over rooms, and mixed-group summaries are means over completed rooms within composition. Cost decompositions are calculated within runs or rooms and then averaged with equal weight. Individual histograms and split-half correlations describe participants or agent seats and are not treatment-effect tests.

Independent fair choices give expected imbalance 0.088 and 0.056 and mean travel time $60+40/\Nagents=62.0$ and 60.8 min for 20 and 50 seats, respectively. The individual-dispersion reference assumes constant independent choices with a common probability equal to the observed mean tip-following share. These are sampling benchmarks, not fitted learning models or tests of independence (SI Appendix, sections S4 and S5).

\subsection*{Use of AI tools}
OpenAI Codex (GPT-6, September 2026) and Anthropic Claude (September 2026) assisted with language editing and manuscript revision, and with writing, debugging and checking analysis and visualization code. The authors are responsible for the scientific content, interpretation, references and final verification of all AI-assisted material. This assistance is distinct from the language-model agents studied experimentally, whose configurations are described above.

\section*{Acknowledgments}
This work was partially supported by JSPS KAKENHI Grant Number JP25H01000.

\section*{Competing interests}
The authors declare no competing interest.

\bibliography{references}

\clearpage
\input{02_si}
\end{document}

%% file: 02_si.tex
\setcounter{page}{1}
\renewcommand{\thepage}{S\arabic{page}}
\setcounter{figure}{0}
\setcounter{table}{0}
\setcounter{equation}{0}
\setcounter{footnote}{0}
\renewcommand{\thefigure}{S\arabic{figure}}
\renewcommand{\thetable}{S\arabic{table}}
\renewcommand{\theHfigure}{S\arabic{figure}}
\renewcommand{\theHtable}{S\arabic{table}}
\renewcommand{\theHequation}{S\arabic{equation}}
\renewcommand{\fbar}{\mathbb{E}[f_t]}
\setlength{\emergencystretch}{3em}
\makeatletter
\global\@topnum\z@
\makeatother
\null
\vskip 2em
\begin{center}
{\LARGE \bfseries Supporting Information for\\ Warned alike, AI agents avoid the less-crowded road while people take it\par}
\vskip 1.5em
{\large
\lineskip .5em
\begin{tabular}[t]{c}
Takahiro Ezaki, Naoto Imura and Katsuhiro Nishinari\\[4pt]\normalsize Corresponding author: Takahiro Ezaki (tkezaki@g.ecc.u-tokyo.ac.jp)
\end{tabular}\par}
\vskip 1em
\end{center}
\par
\vskip 1.5em
\thispagestyle{plain}
\noindent\textbf{This PDF file includes:} Supporting text (sections S1 to S10); Figs.~S1 to S10; Tables~S1 to S14.

\section*{Supporting Text}

\subsection*{S1. Verbatim agent prompt and request parameters}
Every language-model decision was a single stateless call with the following user message (placeholders in braces; \texttt{\{LA\}}/\texttt{\{LB\}} are the road labels in alphabetical order after a seed-dependent mapping to physical roads):

\begin{quote}\small\ttfamily\raggedright
You are commuter \#\{agent\_id\}, one of \{n\_agents\} drivers who all commute every weekday morning from the same suburb to the same business district. Exactly two roads connect them: Route \{LA\} and Route \{LB\}. The two roads are physically identical. Travel time on each road depends only on how many of the \{n\_agents\} commuters pick it that morning: a nearly empty road takes about 20 minutes, a road carrying half of the commuters takes about 60 minutes, and if almost everyone crowds onto one road it takes about 100 minutes. Every driver decides independently before departure and cannot see actual traffic until already committed to a road, so your decision must rely on your own recent experience and on the traffic broadcast, when one is provided.

About you: \{sensitivity\_phrase\} You have been making this commute for a while and you know that all the other drivers face exactly the same choice and similar information every morning.

Your own experience over the most recent days (most recent last):\\
\{history\_block\}

Today's traffic broadcast:\\
\{broadcast\_block\}

It is day \{day\}. Decide which route to take this morning. Think about what travel time you expect on each route, including how the other \{n\_others\} drivers might react to the same information you have.

Respond with ONLY a JSON object in exactly this format, with no other text:\\
\{"route": "\{LA\}", "reason": "one short sentence explaining your choice"\}\\
The "route" value must be "\{LA\}" or "\{LB\}".
\end{quote}

The history block lists up to five lines of the form ``Day $d$: you chose Route $X$ and it took $m$ minutes.'' (in the report-history sensitivity, each line also states which road that day's report indicated). The sensitivity phrase is one of three sentences chosen by the agent's time-sensitivity trait $s \sim U(0.8, 1.2)$: ``You are fairly relaxed about your commute time; a slow day annoys you only mildly.'' ($s < 0.93$); ``You care about your commute time about as much as the average driver.'' ($0.93 \le s \le 1.07$); ``You strongly dislike wasting time in traffic and react quickly when your commute gets slow.'' ($s > 1.07$).

\begin{samepage}\emph{Request parameters.} The client-side request logs for the primary model record:
\begin{quote}\small\ttfamily\raggedright
model = gpt-5.4-mini-2026-03-17\\
temperature = 0.7\\
max\_completion\_tokens = 400\\
reasoning\_effort = none\\
response\_format = \{type: json\_object\}
\end{quote}
\end{samepage}
These settings are present in all 118{,}000 core decisions and all 7{,}800 decisions in the ten 20-seat traces used for individual-level analyses (7{,}000 decisions fall within rounds 6--40, before excluding tie rounds). They are recorded request settings, not an independent provider-side measurement of their implementation; the primary-model results concern a configuration that did not request extended reasoning.

The audit also includes five seed-90 P0 screening runs (14{,}750 recorded decisions), conducted under an earlier frozen plan before the P1 execution freeze. These runs do not enter the analyses reported here, except the invented-label F3 screen, which Table~S4 and Fig.~2G show descriptively; one run (2{,}950 decisions) requested temperature 0 rather than 0.7.

The September cross-provider extension contains 29{,}500 decisions per provider. All Claude requests record \texttt{model = claude-haiku-4-5-20251001}, \texttt{temperature = 0.7} and \texttt{max\_tokens = 400}; the backend does not send a thinking argument. Gemini logs record \texttt{model = gemini-3.5-flash}, \texttt{temperature = 0.7}, \texttt{max\_output\_tokens = 400}, \texttt{response\_mime\_type = application/json} and the \texttt{google-genai} SDK. The backend records successful calls with \texttt{thinking\_budget = 0} for every decision. These records establish the submitted configurations and successful requests, not the providers' internal computation. The complete audit is in \path{data/api_settings_audit.json}. Responses were parsed as JSON with a fallback regular expression; first-attempt parse rates were 1.000 in all but three runs (0.998--0.999).

\emph{Seeds.} A run's seed fixes the round-1 assignment, the traits, the label-to-road mapping and the random draws of the learning baseline. The model requests do not set a provider sampling seed. Paired runs therefore share an initial environment, but not a fixed sequence of model samples. The drift control repeats F3 at seeds 3--5 after four weeks (Table~S3 and Fig.~S3); differences between these runs combine sampling variability and any change over time. Prespecified drift flags were $|\Delta I|>0.05$, $|\Delta M|>0.10$ or a change of classified state relative to the original run at the same seed. None was triggered: observed deviations were $|\Delta I|\le0.012$ and $|\Delta M|\le0.023$.

\subsection*{S2. Verbatim broadcasts}
Let $X$ denote the road that was less crowded in the previous round and $Y$ the other road.
\begin{description}
\item[F0] ``(there is no traffic broadcast today)''
\item[F1] ``Yesterday's travel times: Route A = $t_A$ min, Route B = $t_B$ min.'' (integer minutes, labels in alphabetical order)
\item[F2] ``Traffic advisory: Route $X$ is currently less crowded.''
\item[F3] ``Traffic advisory: Route $X$ is currently less crowded. However, many drivers are expected to see this same information and switch to Route $X$, so Route $X$ may become congested.''
\item[F3, stronger wording (F4)] ``Traffic advisory: Route $X$ is currently less crowded. IMPORTANT: many drivers are expected to see this exact same advisory and switch to Route $X$, so Route $X$ may well end up MORE congested than the other route today. Blindly following this advisory could backfire.''
\item[F3, paraphrase 1] ``Traffic advisory: Route $X$ has lighter traffic right now. Keep in mind that many other drivers receive this same advisory and may move to Route $X$, which could leave it congested.''
\item[F3, paraphrase 2] ``Traffic advisory: at the moment fewer cars are using Route $X$. Because a large number of drivers will read this very message and could shift over to Route $X$, it may not stay clear.''
\item[F1, paraphrases] ``Yesterday's measured travel times were $t_A$ minutes on Route A and $t_B$ minutes on Route B.''; ``Traffic report: driving Route A took $t_A$ minutes yesterday; Route B took $t_B$ minutes.''
\item[F2 + numbers, F3 + numbers] The F2 or F3 message followed, on a new line, by the F1 sentence.
\end{description}
When the previous round's counts were equal, F1 reported equal times and F2/F3 still named a road, chosen by a fixed tie-break of the environment (physical road 0 in every archived trace with a tie and in all 119 previous-round ties in rounds 2--40 of the 24 included mixed rooms). These rounds are excluded from $f_t$ (Text S4). The human studies used the F1 and F3 texts verbatim, with the same label convention; the participant exports retain the advisory text and the indicated road for every round.

\subsection*{S3. Multinomial-logit learning baseline}
The comparator population used in the prespecified L1 contrast consists of $N$ agents that hold an exponentially updated estimate of each road's travel time (updated from their own experience and, under F1, from the reported times), map the F2/F3 tip to a prespecified utility bonus for the indicated road, and choose by multinomial logit over the two estimates. Its parameters were fixed before the core was run and are archived with the code. Because the baseline has no representation of the warning sentence, its F2 and F3 conditions are identical by construction (Table~S1). Under F1 the baseline oscillates ($I = 0.44$, $M = 0.88$); the prespecified interaction contrast compares its F0-to-F1 change with that of the language-model population on the same seeds. This comparison characterizes one prespecified adaptive policy. It does not distinguish among possible explanations for the language model's response to F3, such as interpreting the prediction, trusting the source or avoiding announced crowding, because the comparator does not respond to that sentence.

\subsection*{S4. Outcomes, ties, classifier, prespecified tests and reason coding}
Core and exposure-grid outcomes use rounds 6--60; the matched 20-seat runs and human studies use rounds 6--40. Windows for switched and extended runs are stated with those experiments (Text S6 and the main text). $I = \mathbb{E}_t[|p_t - \tfrac12|]$, where $p_t$ is the share of seats on physical Route A. $M$ is the mean over rounds of the share of seats whose road differs from the previous round; a window beginning at round 6 includes the transition from round 5 to 6. Mean travel time averages the round's mean realized travel time over every round of the stated window.

\emph{Tip-following share.} $f_t$ is the share choosing the road with the smaller count in round $t-1$. It is undefined when the previous counts were equal. Throughout this supplement, a ``tie round'' means a round with equal counts in the preceding round. These rounds are excluded from $\fbar$, $\mathrm{Var}(f_t)$ and individual tip-following tendencies in every data set, and retained in $I$, $M$ and travel time. No such rounds occurred in the F0 and F3 core runs, the 100-round extension, the 20-seat F3 runs or the $K = 15$ mixed rooms. They occurred in 30 of the 103 runs with round-level data (at most six per run; 13 in total under F1 and 25 under F2 in the core), in 1--9 of the 35 analyzed rounds of each all-human room (Table~S5), and in 3--11 rounds of the $K = 5$ and $K = 10$ mixed rooms, where near-balanced use makes a 10:10 count common.

\emph{Classifier and tests.} The descriptive frozen classifier requires $I \ge 0.25$, $M \le 0.20$ and at least ten consecutive rounds with the same majority road. After checking this criterion, the archived classifier assigns ``oscillation'' when $I \ge 0.25$, $M \ge 0.60$ and the lag-one correlation of the physical route share is at most $-0.30$; ``near equilibrium'' when $I \le 0.15$; and ``partial'' otherwise. These labels summarize trajectories and are not equilibrium tests. The two prespecified claims of the core were: L1, the interaction $((F0 - F1)_{\mathrm{LLM}} - (F0 - F1)_{\mathrm{MNL}})$ is positive for both $I$ and $M$; L2, $F3 - F2$ is positive for $I$ and negative for $M$. Each component was tested by an exact two-sided sign test on the ten paired seeds (all ten differences in the required direction give $p = 2^{-9} = 0.00195$); a claim's $p$ is the maximum of its two component $p$ values, and the two claims were Holm-adjusted, giving $p = 0.0039$ for both. The claims, the tests, the classifier thresholds, the model and temperature, the analysis window, the prompt and message texts and the confirmatory seeds (3--12) were fixed in a release manifest with SHA-256 hashes of 29 source files, at commit \texttt{31300fb} (20 August 2026, 10:08 JST). The first core request followed at 10:10 JST. Earlier that day, a screening stage of five runs at seed 90 used the preceding code revision with the same thresholds. After it, the written descriptions of L1 and L2 were corrected to match their executable definitions; the executable contrasts, tests and thresholds were unchanged. No confirmatory seed had been run. The registered mixed-group analysis is reported in Text S8; other comparisons are descriptive.

\emph{Individual tendencies.} An individual's tendency is its share of tie-free rounds 6--40 on the tip road. The number of defined rounds is 26--34 in the all-human rooms, 32--35 in the 20-seat GPT runs and 24--35 in the mixed rooms. The reference SD, $\sqrt{q_{\ast} (1 - q_{\ast})/n_{\ast}}$, approximates the dispersion of identical independent choosers at the observed mean tendency $q_{\ast}$, using the mean number of defined rounds $n_{\ast}$. It is a sampling benchmark, not a fitted model of individual differences. Split-half correlation is the Pearson correlation, across individuals, between tendencies over rounds 6--22 and 23--40. Values are: humans under F1, SD 0.25 (reference 0.09), $r = 0.73$; humans under F3, SD 0.27 (0.09), $r = 0.81$; GPT under F1, SD 0.09 (0.08), $r = 0.14$; GPT under F3, SD 0.03 (0.03), $r = 0.23$. In mixed rooms, human-seat SDs are 0.24, 0.16 and 0.13 and correlations are 0.66, 0.58 and 0.75 at $K = 5$, 10 and 15. These pooled descriptive correlations retain shared room histories and are not estimates of context-independent traits. The GPT values come from five 20-seat runs per condition, seeds 3--7 (\texttt{results/matched/m20\_\{F1,F3\}\_s\{3..7\}\_trace.jsonl}; 100 agents per condition); the all-human values come from twelve benchmark rooms (120 participants per condition).

\emph{Conditional switching.} For Fig.~3E, each seat's switching frequency is calculated separately after a slow ($>60$ min) or fast ($<60$ min) round, using transitions from rounds 6--39 to rounds 7--40. Seats without an eligible transition for a category are omitted from that category's mean. Reported probabilities are means of these seat-level frequencies, rather than pooled frequencies over all eligible transitions. Travel times of exactly 60 min are excluded from this conditioning. This diagnostic uses one fewer transition than the primary mobility outcome $M$, which also includes round 5 to 6.

\emph{Reason coding.} The visible one-sentence reasons from seed 3 of the 50-agent core (rounds 6--60; 2{,}750 decisions per condition) were lower-cased and classified by fixed regular expressions. Code A requires a reference to other drivers followed by a switching or crowding term; code B requires both a slowness term and a self- or experience-reference. The exact expressions are in \path{build/traces.py} in the reproducibility archive (Text S10). Shares for A, B and both were 0.76, 0.25 and 0.21 under F1; 0.61, 0.07 and 0.04 under F2; and 0.89, 0.55 and 0.50 under F3. These coarse codes describe text, not internal reasoning; code B can also capture reported road slowness under F1.

\subsection*{S5. The cost identity, tie rounds, the mixing rule and the bounds}
\emph{Identity.} With $n$ of $N$ commuters on a road, that road takes $20 + 80\,n/N$ minutes, so the round's mean travel time is $\frac{1}{N}\big[n\,(20 + 80\,n/N) + (N-n)\,(20 + 80\,(N-n)/N)\big] = 20 + 80\,[p^2 + (1-p)^2] = 60 + 160\,(p - \tfrac12)^2$ with $p = n/N$. Because the share on the previously less-crowded road is $f_t$ and the share on the other road is $1 - f_t$, $(p_t - \tfrac12)^2 = (f_t - \tfrac12)^2$, and averaging over the rounds on which $f_t$ is defined gives Eq.~1 of the main text with the population variance of $f_t$. The relation is an identity of the cost function; it holds for any population and any composition of a room, and it is exact on the tie-free rounds (largest numerical residual $5 \times 10^{-14}$ min). Equation~1 and both axes of Fig.~3A use the same tie-free rounds. The primary travel-time summaries elsewhere use all rounds of the analysis window, so they can differ from the tie-free mean: this difference is zero for runs without ties, below 0.5 min in 95 of the 103 runs, below 1 min in 100, and at most 1.65 min (20-agent F1, seed 5, two tie rounds); it is below 0.33 min in every all-human and mixed room. The lean and volatility components in Fig.~3C are $160(\fbar - \tfrac12)^2$ and $160\,\mathrm{Var}(f_t)$ over tie-free rounds.

\emph{Linear mixing.} For the exposure grid, the mixing rule takes the pure-population leans $\mathbb{E}_{F1}[f_t] = 0.695$ and $\mathbb{E}_{F3}[f_t] = 0.038$ from the core, forms $\mathbb{E}_{\mathrm{mix}}[f_t] = (1 - x)\,\mathbb{E}_{F1}[f_t] + x\,\mathbb{E}_{F3}[f_t]$ for a warned fraction $x$, and predicts $I \approx |\mathbb{E}_{\mathrm{mix}}[f_t] - \tfrac12|$ and $\mathbb{E}[T_t] \approx 60 + 160\,(\mathbb{E}_{\mathrm{mix}}[f_t] - \tfrac12)^2$. This approximation assumes that the pure-population leans persist under mixing and neglects temporal variation. Its minimum lies at $x = (\mathbb{E}_{F1}[f_t] - \tfrac12)/(\mathbb{E}_{F1}[f_t] - \mathbb{E}_{F3}[f_t]) = 0.30$. Table~S7 also gives a homogeneous independent-choice benchmark at the same mean: $I_{\mathrm{ind}} = \mathbb{E}[|X/N - \tfrac12|]$ and $T_{\mathrm{ind}} = 60 + 160[(\mathbb{E}_{\mathrm{mix}}[f_t] - \tfrac12)^2 + \mathbb{E}_{\mathrm{mix}}[f_t](1 - \mathbb{E}_{\mathrm{mix}}[f_t])/N]$ for $X \sim \operatorname{Binomial}(N, \mathbb{E}_{\mathrm{mix}}[f_t])$. This benchmark gives every seat the same choice probability; it does not model the distinct policies of the two exposure groups. Both benchmarks approximate the U-shape and the location of the minimum but under-predict imbalance near it (5 of 20 warned: 0.03 and 0.09 against 0.14 observed; 12 of 50: 0.04 and 0.06 against 0.13). These discrepancies identify limits of the constant-mean approximations, without isolating temporal adaptation or dependence as their cause. The same mixing rule with the all-human lean $\mathbb{E}_{\mathrm{h}}[f_t] = 0.51$ and the 20-seat agent lean $\mathbb{E}_{\mathrm{a}}[f_t] = 0.03$ gives the dashed line of Fig.~5A. The slight turn of the Fig.~5A dashed line near $K=0$ follows from the all-human mean being just above one half: adding a small agent fraction first brings the mixture to one half, before it falls below it. The plotted continuous reference includes this exact crossing; it does not imply that fractional agents were tested.

\emph{Bounds and reconstruction in mixed rooms.} If all $K$ agents sit on one road in a round, the count on that road is at least $K$, so $|p_t - \tfrac12| \ge \max(0, K/20 - \tfrac12)$ whatever the humans do; this conditional bound is the solid line of Fig.~5A. The bound given the agents' actual choices is obtained as follows. If $a_t$ agents sit on Route A, the humans can move the Route-A count to any integer in $[a_t, a_t + 20 - K]$. The smallest attainable $|p_t - \tfrac12|$ is zero if this interval contains 10, and otherwise the distance of its nearer end from 10, divided by 20. Its mean over rounds 6--40 is 0 at $K = 5$ and $K = 10$ and 0.238 at $K = 15$ (room range 0.226--0.246), against observed 0.241--0.293. All agents occupied one road in 63--91\% of rounds at $K = 15$ (mean 80\%), in 80--97\% at $K = 10$, and in every round at $K = 5$.

On a defined-tip round, if all agents avoid that road and $K \ge 10$, its count is $(20 - K)\,f_{\mathrm{h},t} \le 10$. Thus $|p_t - \tfrac12| = \tfrac12 - (20 - K)\,f_{\mathrm{h},t}/20$, regardless of correlations among human choices. Averaging over the same defined-tip rounds gives $\mathbb{E}_{\mathcal R}[|p_t-\tfrac12|] = \tfrac12 - (20-K)\,\mathbb{E}_{\mathrm{h}}[f_t]/20$. Substituting the observed human tendencies gives 0.084 at $K = 10$ and 0.272 at $K = 15$, compared with the all-round observed means 0.085 and 0.259. These comparisons also reflect the agents' occasional tip choices and, at $K = 10$, the different treatment of tie rounds. At $K = 5$ the tip-road count can exceed 10; an independent-choice calculation with $X \sim \operatorname{Binomial}(15, \mathbb{E}_{\mathrm{h}}[f_t])$ gives 0.073, versus 0.069 observed over all rounds. These calculations describe consistency with mean behavior; they do not test independence among humans.

\emph{Near-threshold runs.} Three Claude F3 runs had $I>0.40$ but $M=0.205$--$0.209$, just above the frozen criterion of $M\le0.20$ (Table~S3). The continuous outcomes therefore remain the primary description of their concentration.

\subsection*{S6. Switch experiments round by round}
Table~S9 lists, for each seed, the mean travel time in round 30 (last round under the original message), the tip-following share, imbalance, switching fraction and mean travel time in round 31 (first round under the new message), the travel time in each of rounds 32--35, and the mean over the prespecified post-intervention window (rounds 36--50). Under F3$\rightarrow$F1, the first response was near-unanimous movement onto the road that had been almost empty ($f_{31} = 0.96$--$1.00$, $I_{31} = 0.46$--$0.50$, $M_{31} = 0.86$--$0.98$, 94--100 min). The persistent majority on the original road ended immediately, but congestion remained high in that round. From round 32, oscillations decreased in amplitude; outcomes in rounds 36--50 were similar to stationary F1 runs over the same window (67--74 min versus 70--74 min in the ten core F1 runs; $I$ 0.19--0.25 versus a median of 0.23). Under F1$\rightarrow$F3, no agent chose the tipped road in round 31 in any seed ($f_{31} = 0$, 100 min), and rounds 36--50 averaged 91--96 min. F3$\rightarrow$F1 replaces the entire warning-bearing advisory with numerical reports. It therefore estimates the response to that message replacement, rather than to deletion of the warning sentence alone.

\subsection*{S7. Human study}
\emph{Participant characteristics and repeat participation.} The analyzed samples contained 240 people in each study. Repeat participation was prohibited, including participation in earlier project experiments. Local research-identifier checks confirmed 240 distinct participants within each analyzed sample and no overlap between them (480 distinct people). The table reports optional age-group and gender responses as counts and percentages of all 240 participants per study; ``prefer not to say'' and missing responses are separate. Demographic responses did not determine analysis inclusion. Only aggregate demographic counts are released (\path{data/participant_demographics.json}).

\input{participant_demographics}

\emph{Recruitment and procedure.} Participants were recruited through Prolific (fluent in English, desktop device, approval rate at least 95\%) into 20-person rooms with an anonymous 180-s preparation lobby and a 30-s presence confirmation. Each choice round consisted of a 9-s advisory period, a decision window of 15 s in rounds 2--5 or 10 s in rounds 6--40, and 3 s of own-outcome feedback. The five most recent own outcomes remained visible. The twelve analyzed rooms formed three blocks of four, with two F1 and two F3 rooms per block assigned by a committed within-block randomization. Each room began with 20 participants and contributed 800 seat-round records, including any carried choices.

\emph{Participant flow and registration.} The plan was registered on 25 August 2026. A timing pilot (40 participants) and technical retests (four participants) were excluded. The twelve benchmark rooms completed on 1--6 September after retries of rooms that had not formed; no room was stopped or replaced after round 1 began, and every formed room is analyzed. Amendments revised lobby, compensation and recruitment procedures. The amendment authorizing the last B1 retry disclosed that the three completed B1 rooms had already been inspected descriptively. Detailed amendments are in the OSF registration (\url{https://osf.io/whsyu}; currently under embargo and not publicly accessible). Waiting compensation was settled for all unseated entrants, although the operational records do not support a complete count of entrants to failed attempts.

\emph{Timeouts and payment.} A timeout carried the previous road forward. After three consecutive timeouts, the last road was carried forward permanently, including after reconnection. The primary room outcomes retain these choices. The registered no-timeout sensitivity excludes records flagged as a current timeout or dropout carry-forward, recalculates road shares over the remaining seats, and averages over the same outcome window. It gives $I = 0.089$ and $M = 0.474$ for F3 rooms, versus 0.085 and 0.456 in the primary summaries. Pay consisted of a base payment of GBP 4.50 plus a bonus of GBP $2\,(100 - \mathbb{E}_t[t_i(t)])/80$, bounded between GBP 0 and GBP 2, using the participant's mean travel time over all 40 rounds. The registered per-protocol sensitivity excludes rooms with at least two permanent dropouts first flagged by round 30; ordinary dropout never invalidates the primary record. One F3 room met this criterion. Retaining five F3 rooms gives mean $I=0.0843$, $M=0.4569$ and travel time 61.75 min; all six F1 rooms are retained ($I=0.0910$, $M=0.4298$, 61.98 min). As registered, these are descriptive retained-room summaries, without a replacement significance test. The included all-human rooms were collected on 1--6 September 2026.

\emph{Questionnaire and consent.} The final questionnaire asked how much the traffic report influenced choices (not at all / slightly / moderately / strongly), how the participant usually responded (took the faster-looking road / took the other road expecting others to switch / ignored it / it depended), and whether the rules were understood (yes / mostly / no). Counts are in Table~S6. Informed consent was obtained under approval E26ALS0545 from the Ethics Review Committee of the Research Center for Advanced Science and Technology, The University of Tokyo. The benchmark disclosure stated that all players were human; the mixed-study disclosure stated that the group included AI-controlled players (Text S8).

\emph{Descriptive message contrast.} The registered estimator averages the difference between the two F3 and two F1 rooms within each block, then averages these differences over the three blocks. Two-sided randomization tests enumerate all $\binom{4}{2}^3 = 216$ within-block assignments; intervals invert a constant-shift test. The estimates are $I$, $-0.005$ ($p = 0.28$; 95\% interval $-0.016$ to $+0.006$; block differences $+0.001$, $-0.014$, $-0.004$), and $M$, $+0.026$ ($p = 0.31$; $-0.037$ to $+0.088$). Applying the same descriptive procedure to mean travel time gives $-0.22$ min ($p = 0.10$; $-0.53$ to $+0.07$). These comparisons carry no confirmatory alpha; non-rejection does not establish equivalence.

\emph{Instructions and questionnaire.} Participants were told that their bonus depended only on their own mean game travel time, not their rank or the group's performance, and could not reduce the base payment. They acknowledged the congestion rule, bonus rule and timeout handling before entering the waiting room. Unlike the agent prompt, the human instructions did not ask them to anticipate others' reactions. Verbatim instructions and questionnaire items are retained in \path{manuscript/paper/tex/human_instructions.tex} and \path{manuscript/paper/tex/questionnaire.tex} in the reproducibility archive (Text S10).

\subsection*{S8. Mixed-group experiment: completion, analysis and limitations}
\emph{Design and completion in brief.} Before mixed-group recruitment began, the 14 September 2026 OSF amendment fixed a target of 24 rooms: eight comparison blocks, each containing three 20-seat rooms, one with 5, one with 10 and one with 15 AI agents ($K=5,10,15$), all under F3. A block is therefore a set of three rooms to be compared, not necessarily a single collection time. This amendment replaced an earlier design that also included new all-human rooms; the existing all-human F3 data remained a contextual reference only. The definitions of $I$ and $M$ and the primary analysis were fixed before mixed-group data collection.

The initial collection on 14--15 September completed 18 of the 24 planned rooms. Two $K=5$ rooms stopped because of a server fault, and four $K=15$ rooms never started because they did not form. Following repair, amendments made before the respective further attempts authorized replacing the two technically invalid rooms and completing the four unstarted rooms. Four rooms were completed on 16 September and the last two on 17 September. These attempts filled the original 24-room design; they did not add new analysis blocks or enlarge the target sample. All 18 initially completed rooms were retained. The two aborted attempts were excluded on technical grounds, kept separately and never pooled with replacement data; their 30 participants received the specified partial payment. The final sample comprised 24 complete 40-round rooms, 240 distinct humans and 240 agent seats.

The authors had seen earlier behavioral results when making the completion amendments. These amendments were prospective for the remaining attempts, not for the full dataset, and explicitly permitted exceptions to the original prohibition on repeating a started room. The primary outcome definitions and eight-block analysis remained unchanged. The detailed protocol revisions and amendment history are provided in the OSF registration (\url{https://osf.io/whsyu}; currently under embargo and not publicly accessible); room-level completion records are summarized in Figs.~S7 and S8 and Table~S8.

\emph{Procedures and allocation.} Each agent seat received a separate call to \texttt{gpt-5.4-mini-2026-03-17} using the 20-driver prompt (Text S1), without being told that other seats could be human. Humans were told that AI players were present, but not their number or identities. Prior project participants were excluded, and all seats received the same broadcast. Rooms requiring more humans were filled first ($K=5$, then 10, then 15); membership within a selected room followed a committed randomized ordering of ready participants. A prespecified rule allowed feasible remaining rooms to start when recruitment was insufficient. Thus composition could be associated with arrival time and waiting, limiting causal interpretation. A valid room completed all 40 rounds. The analysis retained roads automatically carried forward after timeouts or dropout; technical failures were handled separately as described above.

\emph{What the primary analysis compares.} In each block $b$, we compare imbalance in the room with 15 agents, $I_{b,15}$, with that in the room with 5 agents, $I_{b,5}$. The registered block slope is
\[
b_b=(I_{b,15}-I_{b,5})/2.
\]
The difference spans ten additional agents, so dividing by two expresses it per five additional agents. For example, an imbalance difference of 0.20 gives a slope of 0.10. The primary statistic is the mean of these eight slopes. The $K=10$ rooms remain part of the required design and show the intermediate response; they do not enter this endpoint contrast, which is the equally spaced three-level linear contrast. It does not assume that the observed intermediate response lies on a straight line.

\emph{How the test works.} The one-sided test asks whether the average slope is positive. Keeping the eight observed magnitudes fixed, it reverses or retains each slope's sign and calculates the mean for all $2^8=256$ combinations. The $p$ value is the proportion of these means at least as large as the observed mean, including the observed combination. The registered criterion is a positive mean and $p<0.05$. This calculation assumes $b_b=\beta+\varepsilon_b$: block errors are independent and symmetric about zero, although their variances may differ. Under the null of no common positive shift ($\beta=0$), each sign is equally plausible. These are assumptions about variation across blocks; the recruitment procedure does not guarantee them. The test is therefore model-based, rather than an exact test justified solely by randomized composition assignment. Switching $M$ is a supporting outcome tested for a decrease; it cannot substitute for failure of the primary $I$ criterion.

\emph{Why shared recruitment times matter.} An analysis block and a recruitment window are different units. A window is a period in which participants were recruited together. Originally, each of four windows was scheduled to supply two blocks: (1,2), (3,4), (5,6) and (7,8). Blocks recruited together may share participant-pool or timing effects, so their errors need not be independent. The registered sensitivity analysis treats each pair as one unit when reversing signs: both slopes in a pair reverse together, while each of the eight blocks still receives weight $1/8$. There are then only $2^4=16$ combinations. Even the most extreme result has $p=1/16=0.0625$, so this check cannot reach the 5\% threshold.

The later recruitment periods were needed to finish the missing rooms described above, not to create additional comparison blocks. On 16 September, one period supplied pending rooms from blocks 4, 5 and 7, and another from blocks 6 and 8. These periods linked blocks that belonged to different original pairs. An additional sensitivity analysis therefore grouped together all blocks connected by any shared recruitment period, directly or through another block. This gives two groups, $\{1,2\}$ and $\{3,4,5,6,7,8\}$. Completion of blocks 4 and 5 on 17 September does not change those groups. Reversing all slopes within each group together leaves only $2^2=4$ combinations and a minimum $p=0.25$, again with equal weight per original block.

Both grouped checks allow dependence within a group but assume independent groups and joint sign symmetry of each group's errors: reversing the entire error vector must be equally plausible under the null. They do not remove every possible dependence. In particular, the four initial windows occurred during one evening and the following midnight period, so effects shared across the whole night could link even different groups. These checks show how inference changes when fewer independent units are assumed; neither establishes independence nor supplies an alternative criterion for confirmatory success.

\emph{Results and their interpretation.} All eight imbalance slopes were positive, averaging $+0.0951$ per five additional agents. Only one of the 256 sign combinations gave a mean at least this large, so $p=1/256=0.00390625$. The primary criterion is met under the independent-block model. All eight supporting switching slopes were negative (mean $-0.2080$, $p=1/256$). When signs were reversed by the four original recruitment pairs, both outcomes gave $p=0.0625$; grouping all shared-window connections gave $p=0.25$. The direction of the observed differences does not change, but neither dependence check provides significance at 5\%. The evidence from the primary test therefore remains conditional on its independence assumption. Omitting any one block retained both directions (Table~S10).

The registered interval reports the smallest and largest block slopes: $[0.0764,0.1150]$ for $I$ and $[-0.2300,-0.1886]$ for $M$. To interpret this range as an uncertainty interval for a shared central effect, the eight slopes must be independent and have a common median. Under those assumptions, the chance that all eight fall on the same side of that center is at most $2/256$; the range therefore covers it with probability at least $254/256$ (99.2\%). This is not an interval for an arbitrary average of different block effects, nor is it valid without the stated assumptions.

Registered secondary analyses include leave-one-block-out mean slopes, adjacent contrasts and curvature, the no-timeout sensitivity, and the perceived share of automated players compared with $K/20$. The recovered composition estimates are reported below. Carryover and waiting-time sensitivities subtract hypothetical nuisance effects from room outcomes: $Y_{\mathrm{adj}}=Y-\gamma\,\mathrm{carryover}$ for $\gamma \in \{0,\pm0.005,\pm0.01,\pm0.02,\pm0.05\}$ and, separately, $Y_{\mathrm{adj}}=Y-\gamma\,\mathrm{wait}$ for $\gamma \in \{-0.005,0,0.005\}$ per minute. Here carryover refers to a later operational attempt relative to the original recorded attempt, including earlier zero-room attempts, and wait is the room's mean human entry-to-start time. These grids describe sensitivity to hypothetical bias, not identified causal adjustments. Historical pure-human F3 rooms remain outside the mixed-study blocks.

\emph{Perceived automated share: recovery and missingness.} An optional final question asked what percentage of the 20 players, including the respondent, were automated. The 0--100\% slider recorded no answer unless moved. The original exports omitted this field; on 22 September, a read-only, hash-checked recovery retrieved the original submitted responses without changing sample membership. These were original end-of-task answers, not a follow-up survey. Of 240 humans, 238 submitted the questionnaire and 230 supplied an estimate: 116/120, 76/80 and 38/40 at $K=5$, 10 and 15. Unsubmitted drafts were excluded and missing answers were not imputed. Recovery provenance is retained in \path{data/h2_belief_summary.json}.

\emph{Descriptive belief results.} With equal weight per respondent, estimated AI percentages averaged 63.1\%, 67.2\% and 68.9\% for actual percentages of 25\%, 50\% and 75\% (fig.~S9; table~S12). Medians were 65\%, 70\% and 70\%; mean signed errors were $+38.1$, $+17.2$ and $-6.1$ percentage points, and mean absolute errors were 39.4, 20.4 and 17.1 points. Equal weighting of the eight room means per composition gave 63.2\%, 67.4\% and 69.3\%, respectively. As an additional descriptive missing-data bound, assigning every missing estimate any value from 0 to 100\% places the all-participant means in 61.0--64.3\%, 63.9--68.9\% and 65.5--70.5\%; these bounds are not imputations or confidence intervals. Reported composition therefore varied much less than actual composition, and overestimation was largest at $K=5$. This single post-task item does not measure beliefs during play or understanding of the agents' route-choice policy. No belief-based hypothesis test, primary covariate adjustment or exclusion was performed.

\emph{Checks across collection dates.} Four of the eight primary $K=15$ versus $K=5$ comparisons combine rooms collected on different nights. Later $K=15$ rooms had $I=0.241$--$0.260$, overlapping the initial range of $0.244$--$0.293$; the two replacement $K=5$ rooms had $I=0.061$ and 0.069, within the initial range of $0.059$--$0.091$. Every $K=15$ room had higher imbalance than every $K\le10$ room. These descriptive checks show consistent ordering, but do not remove confounding by collection date or participant recruitment. Room-by-room dates and outcomes are in Fig.~S7 and Table~S8.

\emph{Exploratory uncertainty check (not registered).} An equally weighted room-level OLS model with composition as a factor gave the contrast $(I_{15}-I_5)/2=0.0951$. Its conventional standard error was 0.0036; CR1 cluster-robust standard errors were 0.0045 for seven recruitment windows and 0.0042 for eight original blocks. With so few clusters, these standard errors are descriptive checks of uncertainty under alternative covariance assumptions, not evidence that errors are independent or a separate confirmatory test. The registered block-level analysis remains primary.

\emph{Additional registered summaries.} The 24 completed rooms are reported in Fig.~5, Fig.~S7 and Table~S8. Their blockwise $K=15$ minus $K=5$ differences in $I$ are +0.199, +0.230, +0.153, +0.190, +0.171, +0.180, +0.200 and +0.199. Mean adjacent contrasts ($K=10$ minus 5, then $K=15$ minus 10) are $+0.0161$ and $+0.1741$ for $I$ and $-0.2991$ and $-0.1170$ for $M$; mean curvature is $+0.1580$ and $+0.1821$, respectively. Table~S11 reports every registered carryover and waiting-time sensitivity. Carryover is derived from the earliest actually opened window for the original block, including zero-room predecessor records for blocks 1, 2 and 5; a scheduled but unopened window is not an attempt. Across the specified grids the mean $I$ slope remains positive and the mean $M$ slope negative. The no-timeout sensitivity, excluding both current timeouts and dropout carry-forward as defined in Text S7, gives $I = 0.071$, 0.086 and 0.266 at $K = 5$, 10 and 15, versus 0.069, 0.085 and 0.259 with carry-forward retained. Corresponding $M$ means are 0.480, 0.173 and 0.056, versus 0.471, 0.172 and 0.055. For this mobility sensitivity, each retained current choice is compared with that seat's saved previous-round choice; a missing current choice is removed rather than bridging across rounds. No-timeout slopes are $+0.0974$ for $I$ and $-0.2121$ for $M$. No included mixed room had two permanent dropouts first flagged by round 30, an additional descriptive attrition check. The current H2 registration does not specify that threshold as a room-exclusion rule. The audit of dropout thresholds, benchmark collection dates and available questionnaire fields is recorded in \path{data/revision_metadata_audit.json}.

\subsection*{S9. Reasoning and current-model extension}
\emph{Design.} This extension was planned after the preceding results were known and was run on 23 September 2026. Five request settings were each tested under F2 and F3 at seeds 3--7, with 50 agents and 60 rounds (50 runs, 147{,}500 decisions). Three settings used \texttt{gpt-5.4-mini-2026-03-17}: no reasoning at temperature 1.0, low reasoning effort and medium reasoning effort. Two used \texttt{gpt-6-luna}: no reasoning and medium reasoning effort, the model's default. The core runs at the same seeds (no reasoning, temperature 0.7) are the reference. Prompts, broadcasts, histories, first-round draws, the parser and the recovery journal came from the unmodified core code. Before the first call of the formal extension, a plan file fixed the design, the SHA-256 hashes of the code files and a descriptive analysis plan. A five-round F3 pilot at low and medium effort, run earlier that day to estimate token use, was not included in the analyses; its records are archived in \path{results/reasoning_pilot_20260923/}. The plan named two summaries: the number of frozen F3 runs in each setting, shown with the core, and the paired F3 $-$ F2 differences in $I$, $M$ and travel time, with sign counts. No significance criterion was added. The plan named medians; we report means, as for the cross-provider extension. The largest difference between the two summaries is 0.005 for $I$, 0.009 for $M$ and 0.7 min for travel time. Ties occurred only under the tip, in 18 of the 50 runs (at most five per run). All-round and tie-free mean travel times differed by less than 0.5 min in 49 runs and by at most 0.78 min.

\emph{Requests.} Reasoning-enabled requests and all Luna requests omitted temperature; their internal sampling temperature is unknown. Completion caps were 400 tokens without reasoning and 4{,}000 with reasoning, including reasoning tokens. All requests specified JSON-object output. Logs retain the accepted parameters and returned model names; Luna is an undated alias, so future calls may use a different version. Each run saved 2{,}950 distinct decisions. Fourteen decisions were retried after token-cap failures; connection/server errors were also retried. Full attempt logs and token counts are archived with the experiment (see Reproducibility below). The provider reports reasoning-token counts but not reasoning content.

\emph{Temperature.} Without reasoning at temperature 1.0, \texttt{gpt-5.4-mini} reproduced the core pattern at every seed. All five F3 runs were frozen. All five F2 runs were near equilibrium (table~S14). Relative to the core runs of the same seeds, F3 travel time was 2.5 min lower (1.6 to 3.3 min) and switching 0.036 higher. These differences combine the change of temperature with any change since the core runs.

\emph{Reasoning.} With low or medium reasoning, no run met the frozen criterion. The warning still raised imbalance, lowered switching, lowered $\fbar$ and raised travel time in each of the ten pairs (table~S13). Compared with the no-reasoning runs at temperature 1.0, reasoning mainly weakened avoidance under the warning. $\fbar$ was 0.26 at low and 0.21 at medium effort, against 0.06. Switching was 0.42 and 0.36, against 0.12. Travel time was 70.0 and 74.7 min, against 91.5 min. Under the bare tip, low reasoning increased tip-following ($\fbar = 0.69$, against 0.50). Road shares then alternated from round to round (fig.~S10E). Avoidance under the warning was shared across agents but was not complete for any of them. At low and medium effort, all 250 agents per setting took the tip road in more than 10\% of their defined rounds 6--60. The spread of individual tip-following shares (SD 0.044 and 0.041) was smaller than expected for independent choosers with the same mean (0.059 and 0.054). In the frozen runs at temperature 1.0, 219 of 250 agents took the tip road in at most 10\% of these rounds.

\emph{GPT-6 Luna.} Without reasoning, Luna moved toward avoidance under the warning in all five pairs but did not freeze ($\fbar$, 0.35 to 0.19; travel time, 67.8 to 78.4 min). At its default medium reasoning, Luna froze in all ten runs, under the tip as well as under the warning ($\fbar = 0.05$ and 0.07; travel time 92.4 and 89.5 min). In this setting the warning made no consistent difference: it raised $\fbar$ in three pairs, left it unchanged in one and lowered travel time in three.

\emph{Cost components and stated reasons.} The warning raised the average-response cost and lowered the temporal-variation cost in all five pairs at temperature 1.0, at low effort and at medium effort, and in four Luna pairs without reasoning (table~S13). With the codes of text S4 applied to seed 3 (rounds 6--60), reasons under the warning referred to other drivers switching or crowding in 86\% of decisions at temperature 1.0, 85\% at low effort, 81\% at medium effort, 80\% for Luna without reasoning and 83\% for Luna with reasoning, against 89\% in the core. Under the bare tip, the share was 88\% for Luna with reasoning, against 61\% in the core. These codes describe the visible one-sentence reasons only.

\emph{Reproducibility.} \texttt{tools/reasoning\_replication\_20260923.py} ran the experiment. Its plan file and client-side attempt log are stored with the runs in \path{results/reasoning_replication_20260923/}. \texttt{build/reasoning\_extension.py} recomputes every number above and tables S13 and S14. It first checks that it reproduces the core run table and the core reason codes. \texttt{build/figS10.py} draws fig.~S10.

\subsection*{S10. Reproducibility of the analysis}
A private Zenodo draft (reserved DOI: \texttt{10.5281/zenodo.22915642}; not publicly accessible) provides \path{reproducibility_review.zip} and six computational archives, \path{computational_records_part01_of06.zip} through \path{computational_records_part06_of06.zip}. Extract all seven into the same directory. The private \href{https://github.com/tkEzaki/shared-warnings-reproducibility}{GitHub repository} contains the reproducibility materials; The complete Zenodo package is designed to support reproduction without GitHub access. These materials are not publicly accessible at the time of this preprint. Public release is planned upon journal publication; the reserved Zenodo DOI is not yet an active public record. Reuse licenses will be specified at release.

Repository-root \path{REPRODUCING.md} gives dependencies and commands; \path{MANIFEST.json} and \path{COMPUTATIONAL_MANIFEST.json} give file hashes. De-identified round observations and closed-choice responses are in \path{human_data/}; scripts and summary data are in \path{manuscript/paper/build/} and \path{manuscript/paper/data/}. Age and gender are released only as aggregate counts. These materials support figure reconstruction and numerical checks without restricted participant exports or new provider requests. Original-source reconstruction requires restricted exports; pinned source hashes and inclusion rules are recorded in \path{manuscript/paper/data/h2_source_manifest.json}. Registered analysis specifications and an amendment summary are in \path{registration/}; the full amendment history is in OSF (Text S8).

Repository-root \path{VALIDATION.json} records the verification of room-level outcomes, automation estimates, manuscript quantities, figures and reasoning-extension analyses. Paths elsewhere in the supporting text are relative to \path{manuscript/paper/} unless stated otherwise.

\clearpage
\section*{Supplementary Figures}

\begin{figure}[h]
\centering\includegraphics[width=\textwidth]{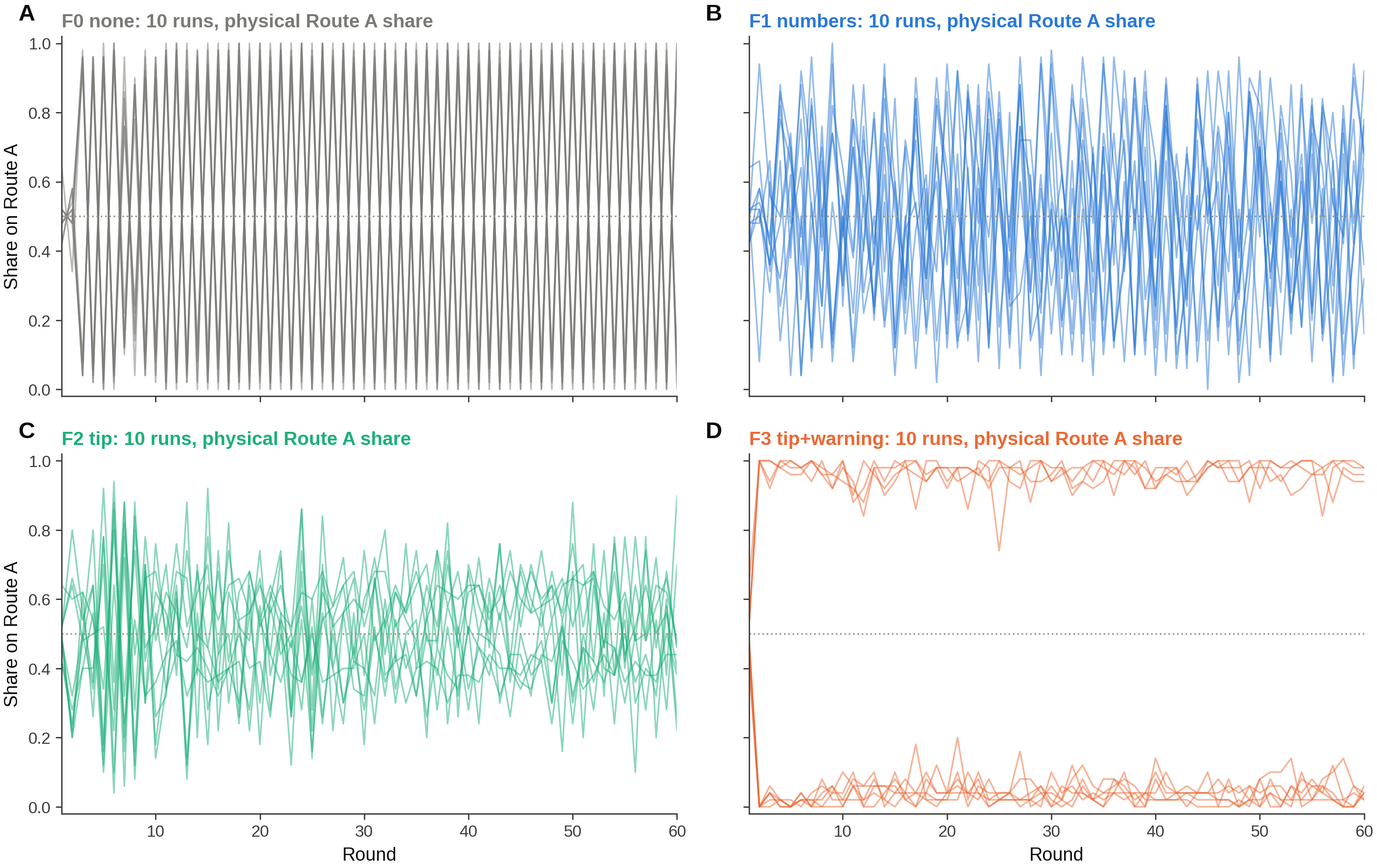}
\caption{\textbf{Physical route shares in the prespecified core.} Share of agents on physical Route A in every run of the four conditions (50 GPT agents, 60 rounds, ten seeds per condition). Road labels are mapped to physical roads with a seed-dependent parity; frozen runs can therefore concentrate on either physical road.}
\end{figure}

\begin{figure}[h]
\centering\includegraphics[width=\textwidth]{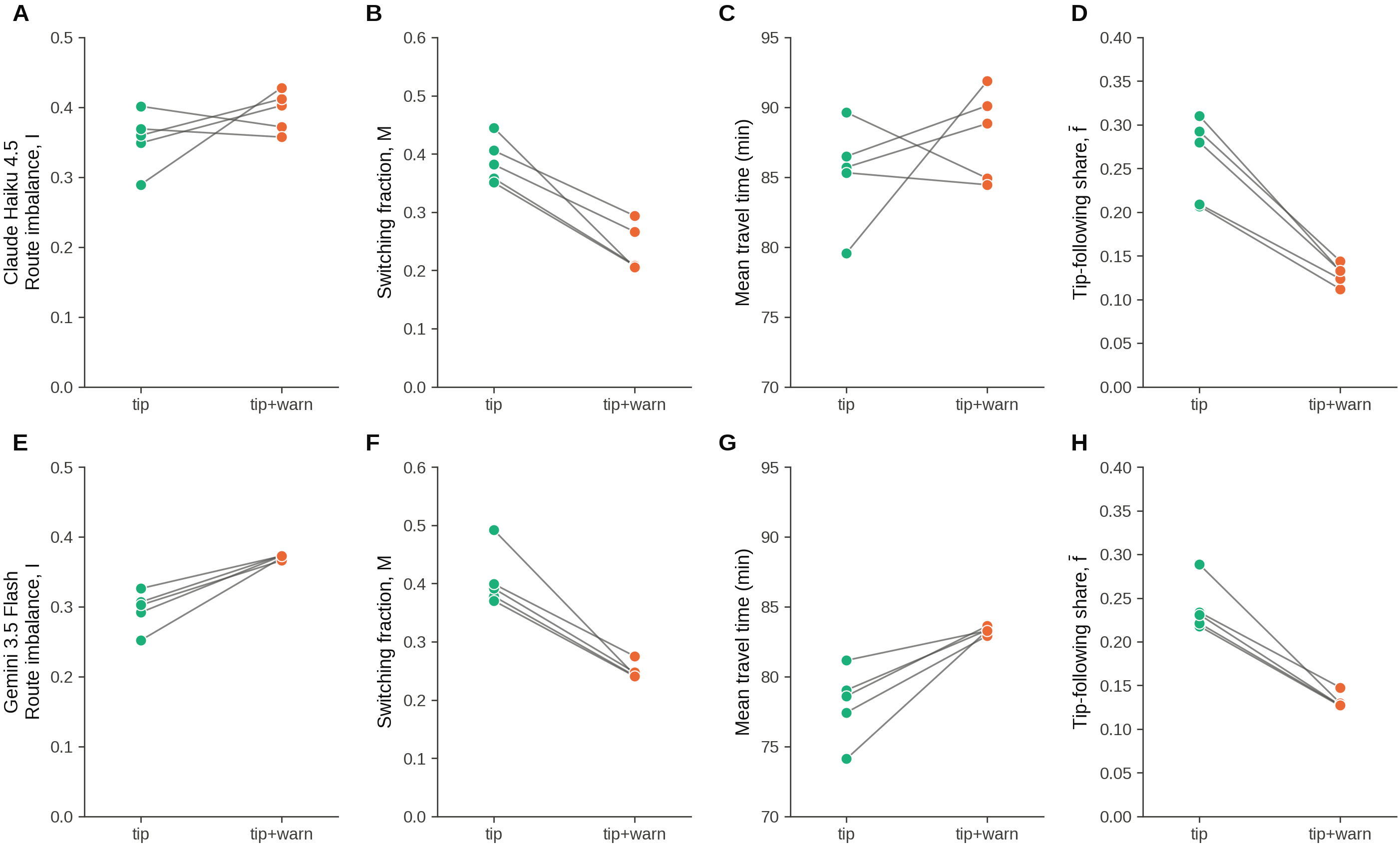}
\caption{\textbf{Cross-provider extension, all paired seeds.} Route imbalance, switching fraction, mean travel time and tip-following share for Claude Haiku 4.5 (\textbf{A} to \textbf{D}) and Gemini 3.5 Flash (\textbf{E} to \textbf{H}) under the tip and under the tip plus warning, five matched seeds each.}
\end{figure}

\begin{figure}[h]
\centering\includegraphics[width=\textwidth]{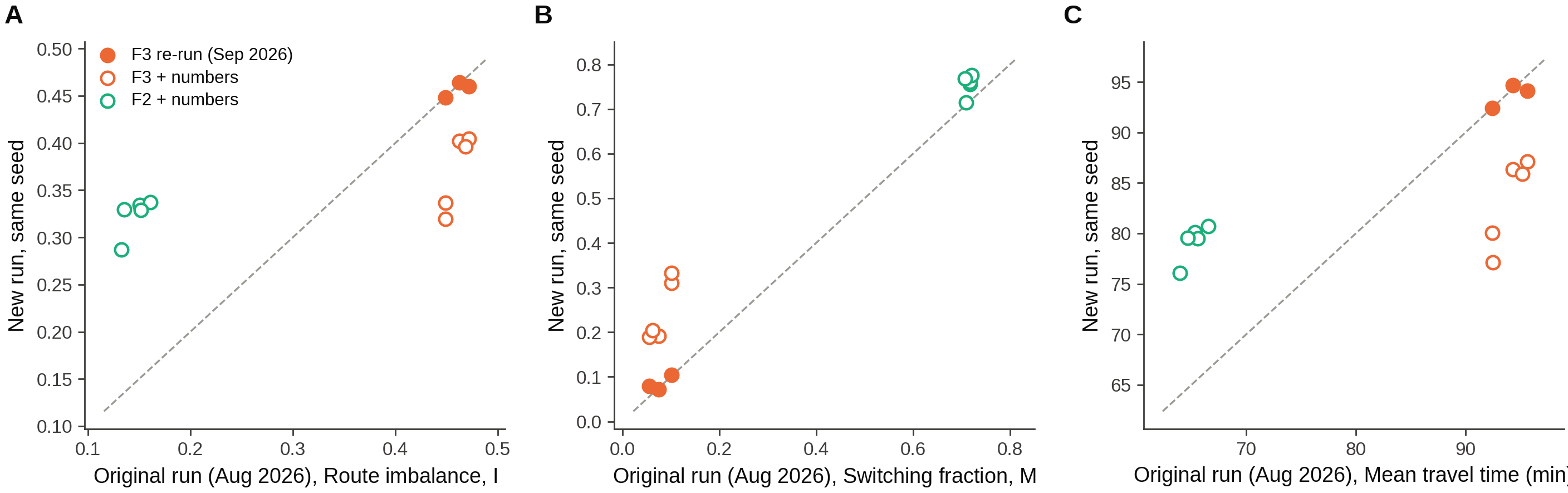}
\caption{\textbf{Drift control and numeric variants by seed.} September 2026 runs compared with August 2026 runs at the same environment seed. Filled points repeat the original F3 condition; hollow points add numerical reports to F3 or F2. The diagonal marks equal outcomes, not identical model samples.}
\end{figure}

\begin{figure}[h]
\centering\includegraphics[width=\textwidth]{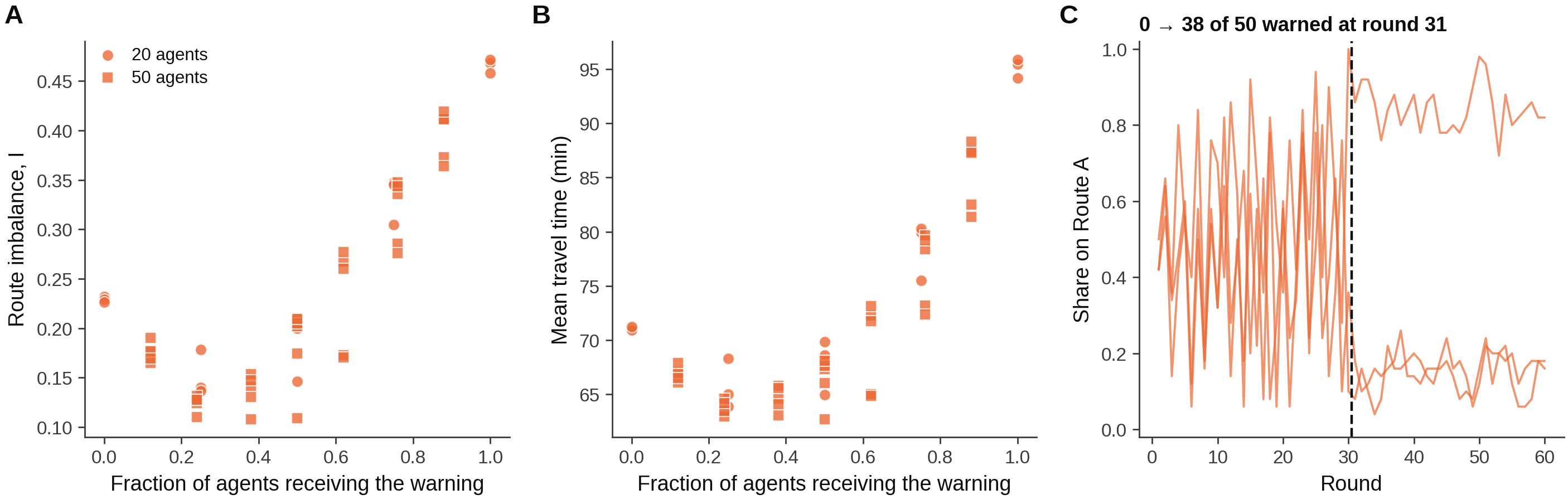}
\caption{\textbf{Exposure grid and exposure switch.} (\textbf{A}, \textbf{B}) Every run of the 20- and 50-agent exposure grids. (\textbf{C}) Physical Route A share in the three runs in which the warning was given to 38 of 50 agents from round 31.}
\end{figure}

\begin{figure}[h]
\centering\includegraphics[width=\textwidth]{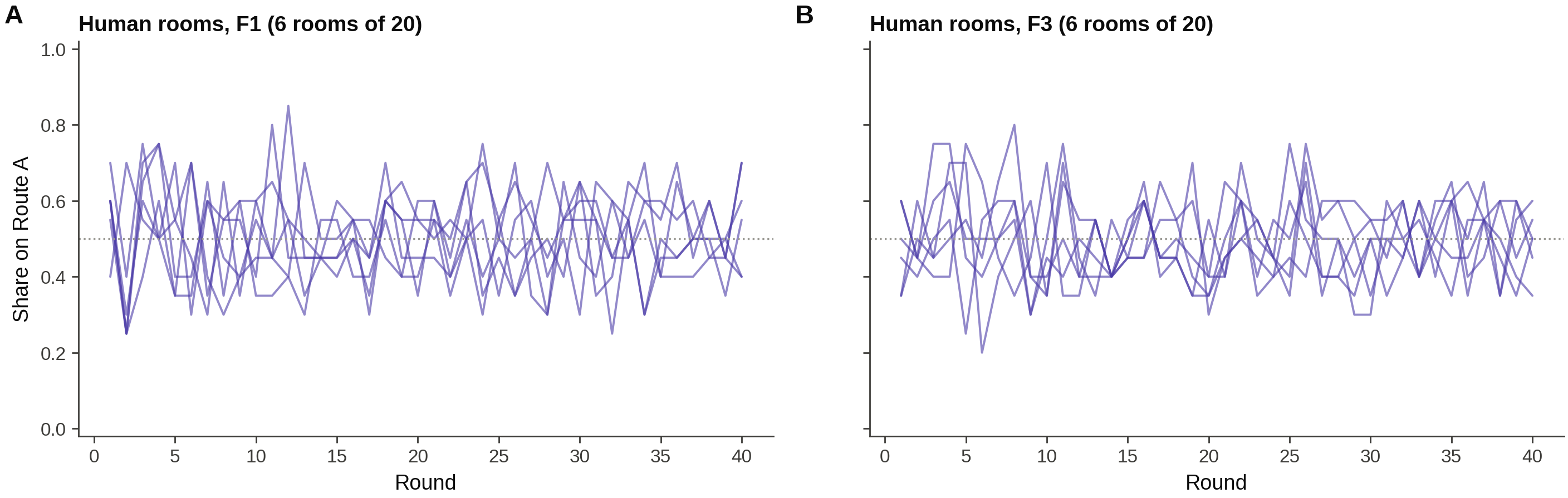}
\caption{\textbf{Human rooms, physical route share.} Share of participants on physical Route A in the six F1 rooms (\textbf{A}) and six F3 rooms (\textbf{B}).}
\end{figure}

\begin{figure}[h]
\centering\includegraphics[width=\textwidth]{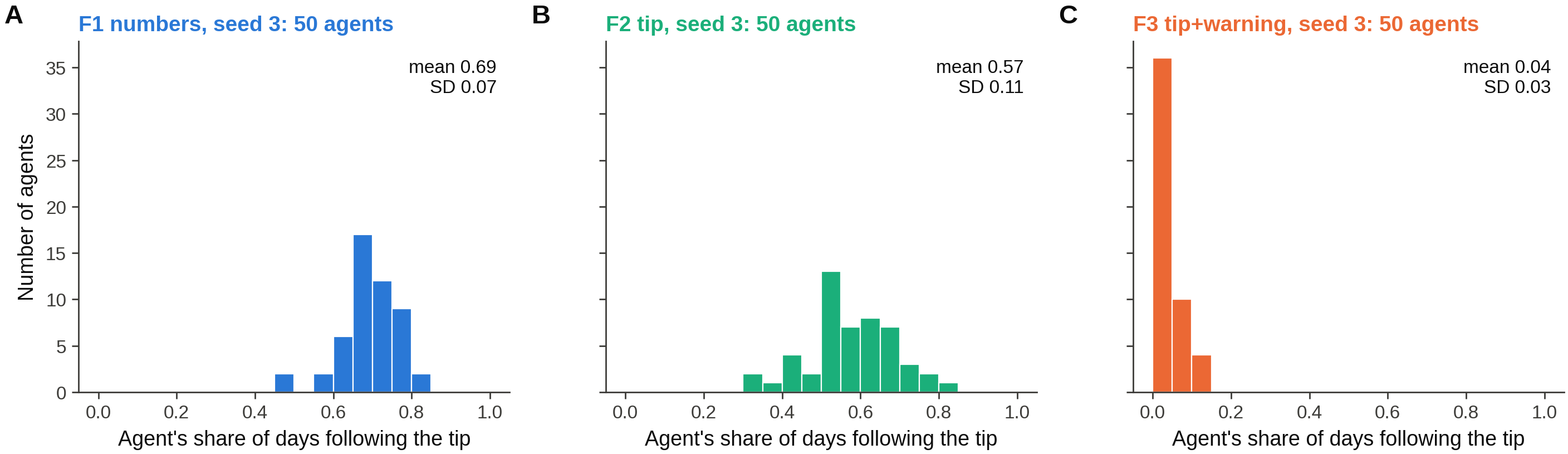}
\caption{\textbf{Individual tip-following tendencies of 50-agent GPT populations.} Each agent's share of tie-free rounds 6--60 on the road that was less crowded in the previous round, seed 3, under numbers (\textbf{A}), the tip (\textbf{B}) and the tip plus warning (\textbf{C}).}
\end{figure}

\begin{figure}[h]
\centering\includegraphics[width=0.95\textwidth]{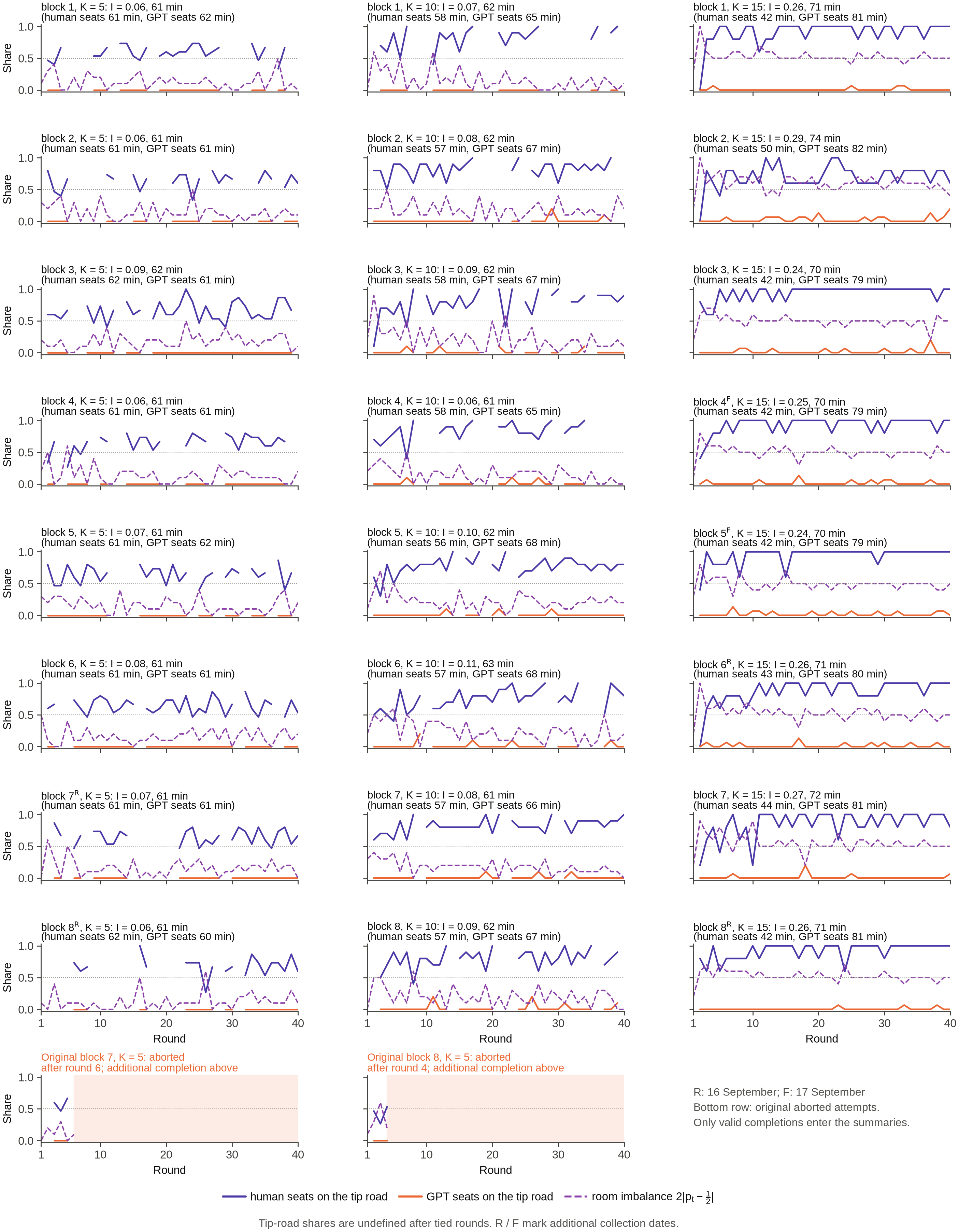}
\caption{\textbf{Every room of the mixed-group experiment.} Columns, $K = 5$, 10 and 15 GPT agents among 20 seats; first eight rows, blocks 1--8; bottom row, the two original aborted attempts. R and F mark valid rooms from the 16 and 17 September collections, respectively. Purple, share of human seats on the tip road; orange, share of GPT seats on it (both undefined on tie rounds); dashed, room imbalance scaled to 1 when everyone is on one road. Titles give $I$ and the mean travel time of the room and of each seat type over rounds 6--40. Aborted rooms show the rounds played before the server fault.}
\end{figure}

\begin{figure}[h]
\centering\includegraphics[width=0.9\textwidth]{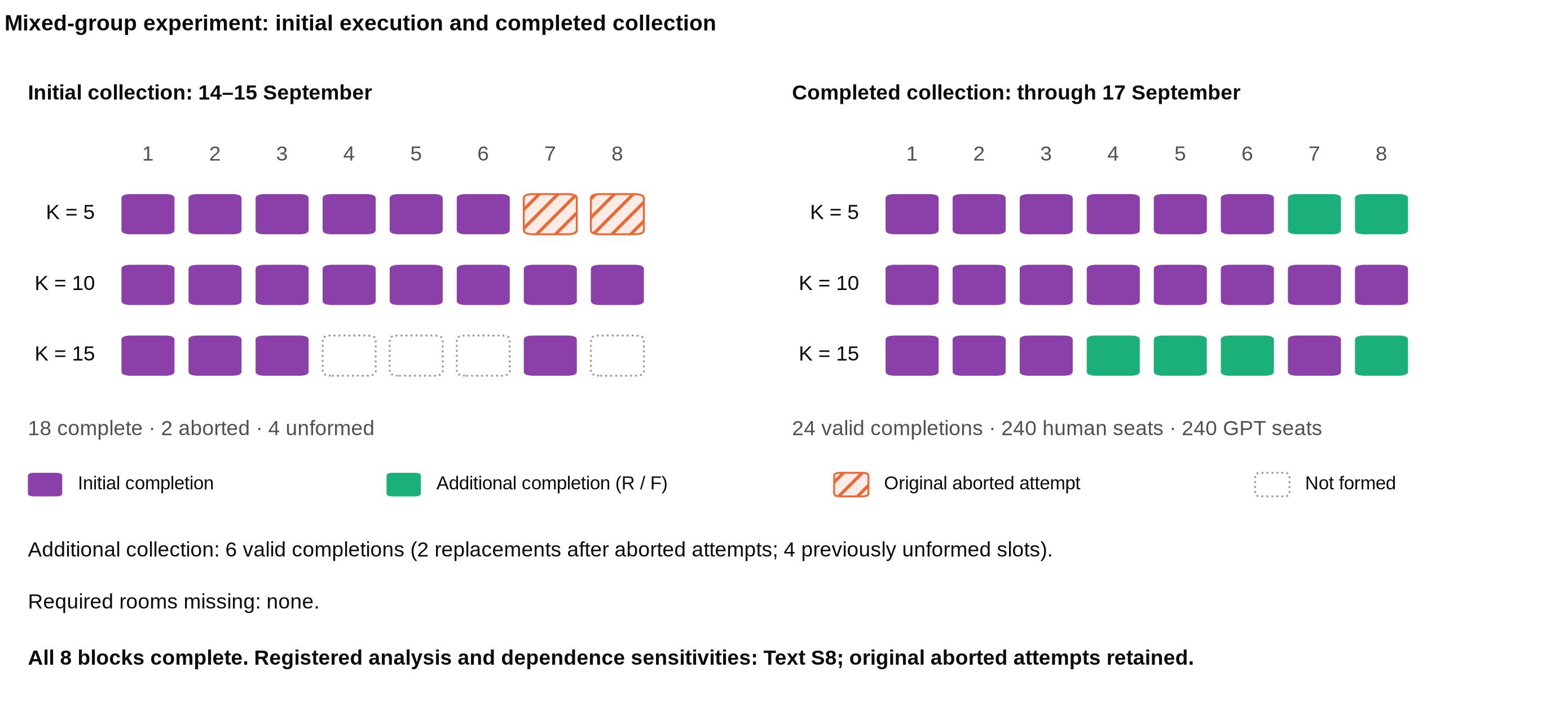}
\caption{\textbf{Design and execution record of the completed mixed-group experiment.} Initial and current status of the 24 registered slots by block and $K$, showing the six valid additional rooms and retaining the two original aborted attempts in the history. All eight blocks are complete (Text S8).}
\end{figure}

\begin{figure}[p]
\centering\includegraphics[width=\textwidth]{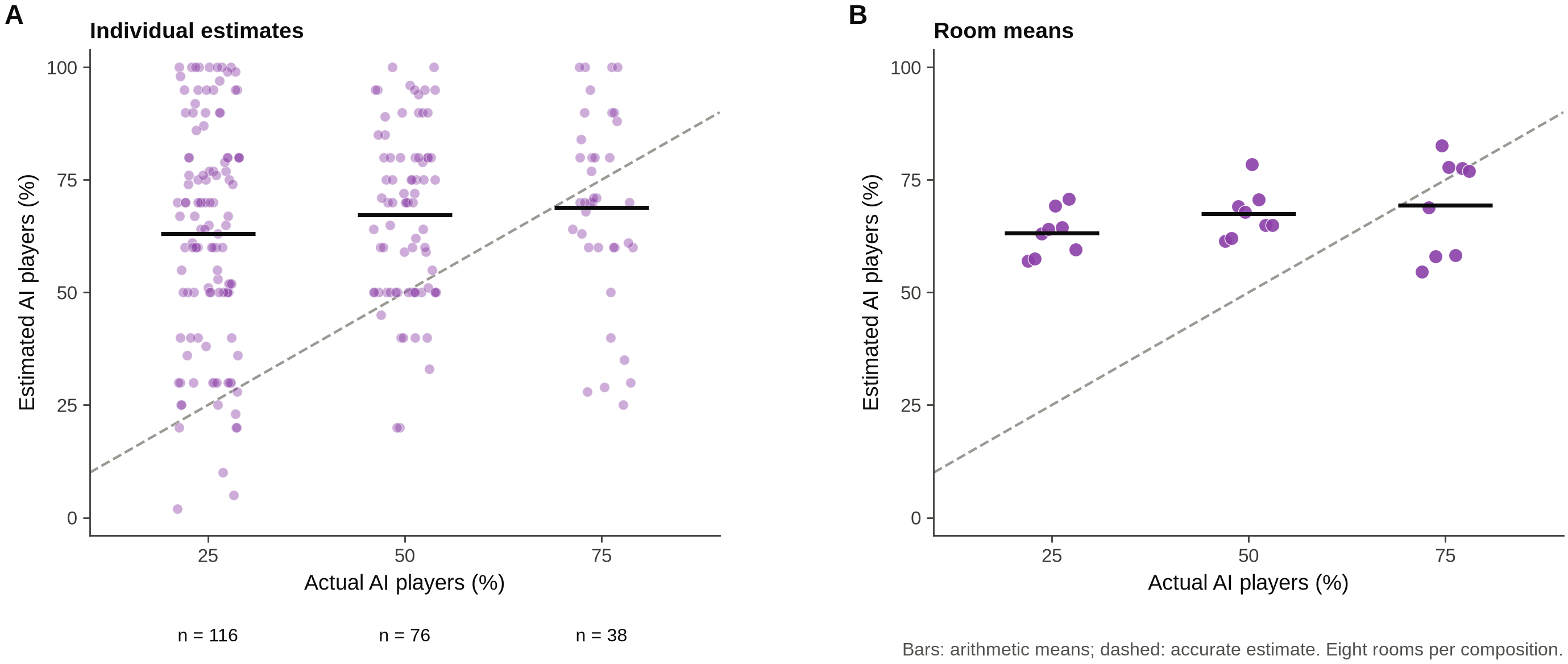}
\caption{\textbf{Post-task estimates of the automated share.} Actual percentages of 25\%, 50\% and 75\% correspond to $K=5$, 10 and 15 of 20 players. (\textbf{A}) Formal numeric answers from 116, 76 and 38 respondents; horizontal bars are respondent means. (\textbf{B}) The eight room means per composition; bars weight rooms equally. Horizontal jitter only separates points. Dashed lines show accurate estimates. Ten missing answers, including one unsubmitted numeric draft, are omitted; all 240 humans remain in the behavioral analysis. These are descriptive distributions, not independent treatment replicates.}
\end{figure}

\begin{figure}[p]
\centering\includegraphics[width=\textwidth]{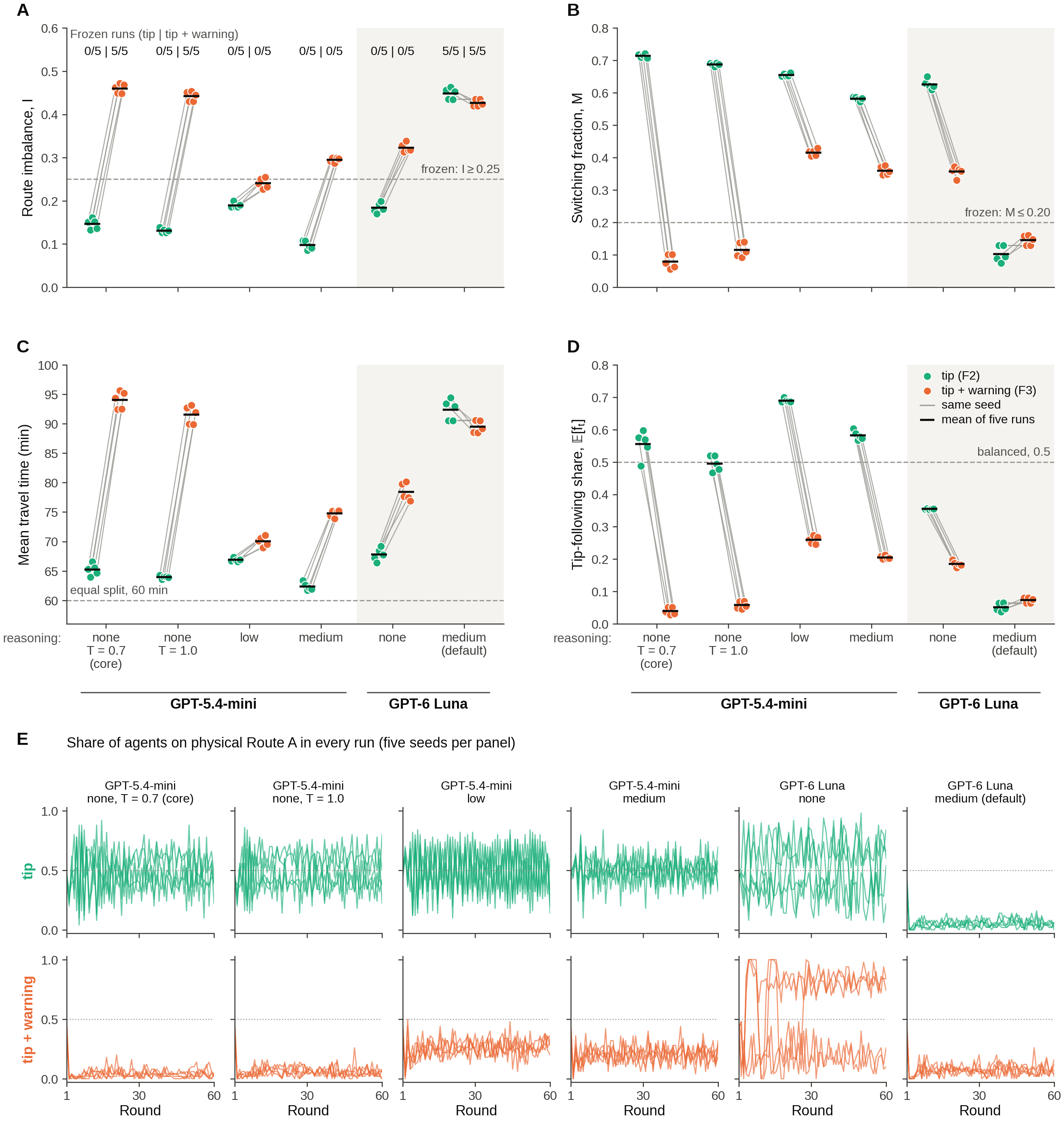}
\caption{\textbf{Reasoning and current-model extension.} (\textbf{A} to \textbf{D}) Route imbalance, switching fraction, mean travel time and tip-following share (rounds 6--60) for each request setting under the tip (F2, green) and under the tip plus warning (F3, orange), five paired seeds each. Gray lines join runs of the same seed; black bars mark means over the five runs. The first setting is the core reference at the same seeds. The numbers at the top of A count runs meeting the frozen criterion, whose thresholds for $I$ and $M$ are dashed (it also requires ten consecutive rounds with the same majority road). Reasoning-enabled requests omitted temperature; its internal numerical value was not independently verified. (\textbf{E}) Share of agents on physical Route A in every run.}
\end{figure}

\clearpage
\section*{Supplementary Tables}

\begin{table}[h]
\centering\small
\caption{\textbf{Condition-level summary.} Medians across runs or rooms of run-mean outcomes; $\fbar$ is the mean across runs of the run-mean tip-following share (tie-free rounds); ``frozen'' counts runs meeting the descriptive classifier. MNL, multinomial-logit learning baseline.}
\input{tableS1}
\end{table}

\begin{table}[h]
\centering\small
\caption{\textbf{All 40 runs of the prespecified core} (50 GPT agents, rounds 6--60; the core was fixed by an execution-time code freeze, not publicly registered).}
\input{tableS2}
\end{table}

\clearpage
\begin{table}[h]
\centering\small
\caption{\textbf{All 33 runs of the September 2026 extension} (50 agents, rounds 6--60). State labels follow the descriptive classifier in Text S4; ``partial'' denotes runs meeting none of its frozen, oscillation or near-equilibrium criteria.}
\input{tableS3}
\end{table}

\begin{table}[h]
\centering\small
\caption{\textbf{Robustness variants} (means across runs; from the audited run table). Rounds 6--40 for 20-seat variants and 6--60 otherwise.}
\input{tableS4}
\end{table}

\clearpage
\begin{table}[h]
\centering\small
\caption{\textbf{Human rooms} (rounds 6--40; carried choices retained). $\fbar$ is the share of participants on the tip road over tie-free rounds; the last column counts the analyzed rounds in which the previous counts were equal.}
\input{tableS5}
\end{table}

\begin{table}[h]
\centering\small
\caption{\textbf{Questionnaire counts.} All-human rooms: 120 participants per message (two F1 and one F3 participant did not answer). Mixed rooms: the 240 participants of the 24 completed rooms (two did not answer).}
\begin{tabular}{p{7.0cm}rrr}
\toprule
Item / response & F1, all human & F3, all human & F3, with agents \\
\midrule
Report influence: not at all / slightly / moderately / strongly & 5 / 16 / 50 / 47 & 3 / 20 / 42 / 54 & 9 / 41 / 90 / 98 \\
Usual response: faster-looking road / other road expecting others to switch / ignored / it depended & 35 / 62 / 4 / 17 & 38 / 47 / 10 / 24 & 146 / 39 / 12 / 41 \\
Rules understood: yes / mostly / no & 104 / 12 / 2 & 108 / 11 / 0 & 202 / 33 / 3 \\
\bottomrule
\end{tabular}
\end{table}

\begin{table}[h]
\centering\small
\caption{\textbf{Exposure grids, the linear-mixing rule and the independent-choice benchmark.} Medians across runs of $I$, $M$ and mean travel time; $\mathbb{E}_{\mathrm{mix}}[f_t]$ mixes the pure-population leans of the core with no fitted parameter; $|\mathbb{E}_{\mathrm{mix}}[f_t] - \frac12|$ and $T_{\mathrm{lin}}$ are the linear-mixing predictions and $I_{\mathrm{ind}}$, $T_{\mathrm{ind}}$ the independent-choice benchmark at the same mean (Text S5).}
\input{tableS6}
\end{table}

\clearpage
\begin{table}[h]
\centering\small
\caption{\textbf{Mixed rooms} (rounds 6--40; carried choices retained). Subscripts h and a denote human and GPT seats; $\fbar$ is the share of that seat type on the tip road over defined-tip rounds. ``Frozen'' applies the descriptive classifier to the room. Tie-round ranges are given in Text S4. All 24 slots are complete; the two invalid original attempts are retained separately in the execution record (Figs.~S7 and S8).}
\input{tableS8}
\end{table}

\begin{table}[h]
\centering\small
\caption{\textbf{Switch experiments round by round.} $T_{30}$, mean travel time in the last round under the original message; $f_{31}$, $I_{31}$, $M_{31}$ and $T_{31}$, the first round under the new message; $T_{32}$ to $T_{35}$, the following rounds; $\mathbb{E}[T_t]_{36\text{--}50}$, the prespecified post-intervention window.}
\input{tableS9}
\end{table}

\clearpage
\begin{table}[h]
\centering\small
\caption{\textbf{Registered block slopes and leave-one-block-out estimates.} Slopes are per five additional agents. Leave-one-block-out columns are mean slopes across the other seven blocks, without additional tests. The full-sample means are 0.0951 for $I$ and $-0.2080$ for $M$ (Text S8).}
\input{tableS10}
\end{table}

\begin{table}[h]
\centering\small
\caption{\textbf{Registered nuisance sensitivity grids.} Each row reports the equally weighted mean block slope after subtracting the stated hypothetical nuisance effect from room outcomes. Carryover is binary and wait is measured in minutes. These are descriptive sensitivity calculations, not estimated causal corrections (Text S8).}
\input{tableS11}
\end{table}

\begin{table}[h]
\centering\small
\caption{\textbf{Recovered formal estimates of the automated share.} Each composition comprises eight rooms. Responses gives numeric respondents / all human participants. Means and SDs weight respondents equally; IQR gives the 25th and 75th percentiles. Bias is estimated minus actual percentage; MAE is mean absolute error, both in percentage points (pp). Missingness comprises eight unanswered items among submitted questionnaires and two unsubmitted questionnaires. No missing or draft-only value is imputed.}
\input{tableS12}
\end{table}

\begin{table}[h]
\centering\small
\caption{\textbf{Reasoning and current-model extension, summary} (50 agents, rounds 6--60, seeds 3--7; means over five runs). The first block is the core reference at the same seeds. ``Response'' and ``variation'' are the two terms of Eq.~1, $160(\fbar-\tfrac12)^2$ and $160\,\mathrm{Var}(f_t)$ in minutes, calculated within each run. Rows ``warning $-$ tip'' give mean paired differences. In parentheses is the number of pairs, of five, whose difference has the sign of the core warning effect (higher $I$ and travel time, lower $M$ and $\fbar$).}
\input{tableS13}
\end{table}

\begin{table}[h]
\centering\footnotesize
\caption{\textbf{All 50 runs of the reasoning and current-model extension} (50 agents, rounds 6--60). ``Ties'' counts analysis rounds whose previous counts were equal; these are excluded from $\fbar$. State labels follow the descriptive classifier in Text S4.}
\input{tableS14}
\end{table}

%% file: participant_demographics.tex
\begin{center}\small
\begin{tabular}{lrr}\toprule
Characteristic & All-human & Mixed\\
 & $n=240$ & $n=240$\\\midrule
\multicolumn{3}{l}{\textit{Age, years}}\\
18--24 & 57 (23.8\%) & 44 (18.3\%)\\
25--34 & 90 (37.5\%) & 99 (41.2\%)\\
35--44 & 56 (23.3\%) & 51 (21.2\%)\\
45--54 & 17 (7.1\%) & 28 (11.7\%)\\
55+ & 13 (5.4\%) & 13 (5.4\%)\\
Prefer not to say & 0 (0.0\%) & 2 (0.8\%)\\
Missing & 7 (2.9\%) & 3 (1.2\%)\\
\multicolumn{3}{l}{\textit{Gender}}\\
Female & 84 (35.0\%) & 111 (46.2\%)\\
Male & 147 (61.3\%) & 122 (50.8\%)\\
Other & 0 (0.0\%) & 1 (0.4\%)\\
Prefer not to say & 0 (0.0\%) & 1 (0.4\%)\\
Missing & 9 (3.8\%) & 5 (2.1\%)\\
\bottomrule\end{tabular}
\end{center}

%% file: tableS1.tex
\begin{tabular}{lrrrrrrrl}
\toprule
Population and message & $N$ & rounds & runs & $I$ & $M$ & time (min) & $\mathbb{E}[f_t]$ & frozen \\
\midrule
GPT, F0 no report & 50 & 6--60 & 10 & 0.476 & 0.993 & 96.4 & 0.96 & 0/10 \\
GPT, F1 numbers & 50 & 6--60 & 10 & 0.230 & 0.660 & 71.5 & 0.69 & 0/10 \\
GPT, F2 tip & 50 & 6--60 & 10 & 0.134 & 0.713 & 64.4 & 0.54 & 0/10 \\
GPT, F3 tip + warning & 50 & 6--60 & 10 & 0.465 & 0.070 & 94.8 & 0.04 & 10/10 \\
MNL baseline, F0 & 50 & 6--60 & 10 & 0.063 & 0.496 & 61.0 & -- & 0/10 \\
MNL baseline, F1 & 50 & 6--60 & 10 & 0.437 & 0.881 & 90.7 & -- & 0/10 \\
MNL baseline, F2 & 50 & 6--60 & 10 & 0.130 & 0.510 & 63.5 & -- & 0/10 \\
MNL baseline, F3 & 50 & 6--60 & 10 & 0.130 & 0.510 & 63.5 & -- & 0/10 \\
GPT, F2 tip + numbers & 50 & 6--60 & 5 & 0.329 & 0.761 & 79.6 & 0.82 & 0/5 \\
GPT, F3 tip + warning + numbers & 50 & 6--60 & 5 & 0.396 & 0.204 & 85.9 & 0.13 & 2/5 \\
GPT, F3 repeat (Sep 2026) & 50 & 6--60 & 3 & 0.460 & 0.079 & 94.1 & 0.04 & 3/3 \\
Claude Haiku 4.5, F2 tip & 50 & 6--60 & 5 & 0.360 & 0.382 & 85.7 & 0.26 & 0/5 \\
Claude Haiku 4.5, F3 tip + warning & 50 & 6--60 & 5 & 0.403 & 0.209 & 88.8 & 0.13 & 0/5 \\
Gemini 3.5 Flash, F2 tip & 50 & 6--60 & 5 & 0.303 & 0.392 & 78.6 & 0.24 & 0/5 \\
Gemini 3.5 Flash, F3 tip + warning & 50 & 6--60 & 5 & 0.373 & 0.243 & 83.3 & 0.13 & 0/5 \\
GPT, F1 numbers (20 seats) & 20 & 6--40 & 10 & 0.242 & 0.702 & 72.7 & 0.72 & 0/10 \\
GPT, F3 tip + warning (20 seats) & 20 & 6--40 & 10 & 0.469 & 0.062 & 95.4 & 0.03 & 10/10 \\
Humans, F1 numbers (20 seats) & 20 & 6--40 & 6 & 0.091 & 0.439 & 61.9 & 0.51 & 0/6 \\
Humans, F3 tip + warning (20 seats) & 20 & 6--40 & 6 & 0.084 & 0.442 & 61.7 & 0.51 & 0/6 \\
\bottomrule
\end{tabular}

%% file: tableS2.tex
\begin{tabular}{lrrrrrr}
\toprule
Condition & seed & $I$ & $M$ & time (min) & $\mathbb{E}[f_t]$ & $\mathrm{Var}(f_t)$ \\
\midrule
F0 & 3 & 0.478 & 0.993 & 96.7 & 0.978 & 0.0006 \\
F0 & 4 & 0.433 & 0.993 & 90.1 & 0.933 & 0.0005 \\
F0 & 5 & 0.474 & 0.995 & 96.0 & 0.974 & 0.0003 \\
F0 & 6 & 0.492 & 0.995 & 98.9 & 0.992 & 0.0009 \\
F0 & 7 & 0.487 & 0.990 & 98.3 & 0.987 & 0.0018 \\
F0 & 8 & 0.414 & 0.988 & 87.6 & 0.914 & 0.0011 \\
F0 & 9 & 0.418 & 0.996 & 87.9 & 0.918 & 0.0001 \\
F0 & 10 & 0.441 & 0.987 & 91.3 & 0.941 & 0.0014 \\
F0 & 11 & 0.489 & 0.993 & 98.4 & 0.989 & 0.0009 \\
F0 & 12 & 0.479 & 0.999 & 96.8 & 0.979 & 0.0000 \\
F1 & 3 & 0.228 & 0.640 & 71.8 & 0.690 & 0.0324 \\
F1 & 4 & 0.224 & 0.667 & 70.5 & 0.687 & 0.0301 \\
F1 & 5 & 0.235 & 0.655 & 71.6 & 0.678 & 0.0385 \\
F1 & 6 & 0.239 & 0.660 & 71.4 & 0.685 & 0.0369 \\
F1 & 7 & 0.232 & 0.678 & 71.1 & 0.708 & 0.0262 \\
F1 & 8 & 0.218 & 0.668 & 70.4 & 0.685 & 0.0307 \\
F1 & 9 & 0.243 & 0.671 & 73.3 & 0.733 & 0.0280 \\
F1 & 10 & 0.218 & 0.644 & 71.0 & 0.689 & 0.0333 \\
F1 & 11 & 0.229 & 0.652 & 72.1 & 0.703 & 0.0299 \\
F1 & 12 & 0.238 & 0.659 & 71.9 & 0.692 & 0.0376 \\
F2 & 3 & 0.151 & 0.718 & 65.3 & 0.575 & 0.0273 \\
F2 & 4 & 0.133 & 0.709 & 64.0 & 0.488 & 0.0242 \\
F2 & 5 & 0.161 & 0.717 & 66.6 & 0.597 & 0.0333 \\
F2 & 6 & 0.152 & 0.721 & 65.6 & 0.570 & 0.0296 \\
F2 & 7 & 0.136 & 0.707 & 64.7 & 0.546 & 0.0281 \\
F2 & 8 & 0.120 & 0.687 & 63.5 & 0.504 & 0.0215 \\
F2 & 9 & 0.129 & 0.725 & 64.0 & 0.477 & 0.0251 \\
F2 & 10 & 0.147 & 0.702 & 65.3 & 0.543 & 0.0286 \\
F2 & 11 & 0.121 & 0.694 & 63.5 & 0.513 & 0.0212 \\
F2 & 12 & 0.133 & 0.718 & 64.1 & 0.540 & 0.0243 \\
F3 & 3 & 0.463 & 0.075 & 94.3 & 0.037 & 0.0007 \\
F3 & 4 & 0.449 & 0.101 & 92.5 & 0.051 & 0.0012 \\
F3 & 5 & 0.472 & 0.056 & 95.7 & 0.028 & 0.0005 \\
F3 & 6 & 0.449 & 0.101 & 92.5 & 0.051 & 0.0018 \\
F3 & 7 & 0.468 & 0.063 & 95.2 & 0.032 & 0.0007 \\
F3 & 8 & 0.468 & 0.064 & 95.2 & 0.032 & 0.0010 \\
F3 & 9 & 0.453 & 0.092 & 93.3 & 0.047 & 0.0025 \\
F3 & 10 & 0.468 & 0.065 & 95.2 & 0.032 & 0.0011 \\
F3 & 11 & 0.461 & 0.076 & 94.3 & 0.039 & 0.0012 \\
F3 & 12 & 0.468 & 0.063 & 95.2 & 0.032 & 0.0008 \\
\bottomrule
\end{tabular}

%% file: tableS3.tex
\begin{tabular}{llrrrrrl}
\toprule
Model & message & seed & $I$ & $M$ & time (min) & $\mathbb{E}[f_t]$ & state \\
\midrule
GPT-5.4-mini & tip + numbers & 3 & 0.334 & 0.756 & 80.1 & 0.833 & oscillation \\
GPT-5.4-mini & tip + numbers & 4 & 0.287 & 0.715 & 76.1 & 0.787 & oscillation \\
GPT-5.4-mini & tip + numbers & 5 & 0.337 & 0.761 & 80.7 & 0.837 & oscillation \\
GPT-5.4-mini & tip + numbers & 6 & 0.329 & 0.776 & 79.5 & 0.829 & oscillation \\
GPT-5.4-mini & tip + numbers & 7 & 0.329 & 0.768 & 79.6 & 0.829 & oscillation \\
GPT-5.4-mini & tip + warning + numbers & 3 & 0.402 & 0.192 & 86.4 & 0.098 & frozen \\
GPT-5.4-mini & tip + warning + numbers & 4 & 0.336 & 0.310 & 80.1 & 0.165 & partial \\
GPT-5.4-mini & tip + warning + numbers & 5 & 0.404 & 0.189 & 87.1 & 0.096 & frozen \\
GPT-5.4-mini & tip + warning + numbers & 6 & 0.319 & 0.332 & 77.1 & 0.181 & partial \\
GPT-5.4-mini & tip + warning + numbers & 7 & 0.396 & 0.204 & 85.9 & 0.104 & partial \\
GPT-5.4-mini & tip + warning (repeat) & 3 & 0.464 & 0.072 & 94.7 & 0.036 & frozen \\
GPT-5.4-mini & tip + warning (repeat) & 4 & 0.448 & 0.104 & 92.4 & 0.052 & frozen \\
GPT-5.4-mini & tip + warning (repeat) & 5 & 0.460 & 0.079 & 94.1 & 0.040 & frozen \\
Claude Haiku 4.5 & tip & 3 & 0.349 & 0.358 & 85.7 & 0.207 & partial \\
Claude Haiku 4.5 & tip & 4 & 0.401 & 0.382 & 89.6 & 0.310 & partial \\
Claude Haiku 4.5 & tip & 5 & 0.360 & 0.351 & 86.5 & 0.209 & partial \\
Claude Haiku 4.5 & tip & 6 & 0.369 & 0.406 & 85.3 & 0.292 & partial \\
Claude Haiku 4.5 & tip & 7 & 0.289 & 0.445 & 79.6 & 0.280 & partial \\
Claude Haiku 4.5 & tip + warning & 3 & 0.403 & 0.209 & 88.8 & 0.112 & partial \\
Claude Haiku 4.5 & tip + warning & 4 & 0.372 & 0.267 & 84.9 & 0.133 & partial \\
Claude Haiku 4.5 & tip + warning & 5 & 0.412 & 0.208 & 90.1 & 0.124 & partial \\
Claude Haiku 4.5 & tip + warning & 6 & 0.357 & 0.294 & 84.5 & 0.144 & partial \\
Claude Haiku 4.5 & tip + warning & 7 & 0.428 & 0.205 & 91.9 & 0.133 & partial \\
Gemini 3.5 Flash & tip & 3 & 0.307 & 0.378 & 79.0 & 0.218 & partial \\
Gemini 3.5 Flash & tip & 4 & 0.252 & 0.492 & 74.1 & 0.289 & partial \\
Gemini 3.5 Flash & tip & 5 & 0.292 & 0.392 & 77.4 & 0.221 & partial \\
Gemini 3.5 Flash & tip & 6 & 0.303 & 0.399 & 78.6 & 0.234 & partial \\
Gemini 3.5 Flash & tip & 7 & 0.326 & 0.371 & 81.2 & 0.231 & partial \\
Gemini 3.5 Flash & tip + warning & 3 & 0.373 & 0.242 & 83.3 & 0.127 & partial \\
Gemini 3.5 Flash & tip + warning & 4 & 0.370 & 0.243 & 83.3 & 0.130 & partial \\
Gemini 3.5 Flash & tip + warning & 5 & 0.373 & 0.248 & 82.9 & 0.127 & partial \\
Gemini 3.5 Flash & tip + warning & 6 & 0.366 & 0.275 & 83.6 & 0.148 & partial \\
Gemini 3.5 Flash & tip + warning & 7 & 0.373 & 0.241 & 83.3 & 0.127 & partial \\
\bottomrule
\end{tabular}

%% file: tableS4.tex
\begin{tabular}{lrrrrl}
\toprule
Variant & runs & $I$ & $M$ & time (min) & frozen \\
\midrule
F3, full history (rounds 6--40, 20 seats) & 5 & 0.479 & 0.041 & 96.9 & 5/5 \\
F1 + past tips in history (20 seats) & 5 & 0.247 & 0.492 & 72.6 & 0/5 \\
F3 + past tips in history (20 seats) & 5 & 0.497 & 0.005 & 99.6 & 5/5 \\
F3, stronger wording (50 agents) & 3 & 0.499 & 0.003 & 99.8 & 3/3 \\
Claude/Gemini, F0 (2 seeds each) & 4 & 0.249 & 0.835 & 73.3 & 0/4 \\
Claude/Gemini, F1 (2 seeds each) & 4 & 0.361 & 0.917 & 83.1 & 0/4 \\
Claude/Gemini, F3 (2 seeds each) & 4 & 0.381 & 0.255 & 85.7 & 0/4 \\
GPT, F1 paraphrase 1 & 3 & 0.275 & 0.734 & 74.7 & 0/3 \\
GPT, F3 paraphrase 1 & 3 & 0.468 & 0.064 & 95.3 & 3/3 \\
GPT, F1 paraphrase 2 & 3 & 0.331 & 0.813 & 79.1 & 0/3 \\
GPT, F3 paraphrase 2 & 3 & 0.373 & 0.251 & 83.0 & 0/3 \\
F3, invented road names Maple/Cedar (screen, seed 90) & 1 & 0.443 & 0.111 & 91.6 & 1/1 \\
\bottomrule
\end{tabular}

%% file: tableS5.tex
\begin{tabular}{llrrrrrr}
\toprule
Room & block & message & $I$ & $M$ & time (min) & $\fbar$ & tie rounds \\
\midrule
pure-b1-2 & 1 & F1 & 0.089 & 0.420 & 61.9 & 0.527 & 4 \\
pure-b1-3 & 1 & F1 & 0.086 & 0.467 & 61.8 & 0.508 & 5 \\
pure-b2-roomwise-r1-2 & 2 & F1 & 0.094 & 0.353 & 62.3 & 0.505 & 4 \\
pure-b2-roomwise-r1-4 & 2 & F1 & 0.093 & 0.457 & 62.0 & 0.503 & 5 \\
pure-b3-roomwise-r3-1 & 3 & F1 & 0.081 & 0.397 & 61.7 & 0.524 & 4 \\
pure-b3-roomwise-r3-2 & 3 & F1 & 0.103 & 0.484 & 62.2 & 0.517 & 2 \\
pure-b1-1 & 1 & F3 & 0.084 & 0.497 & 61.7 & 0.509 & 6 \\
pure-b1-r4x-1 & 1 & F3 & 0.091 & 0.451 & 61.8 & 0.512 & 1 \\
pure-b2-roomwise-r1-1 & 2 & F3 & 0.083 & 0.414 & 61.6 & 0.485 & 5 \\
pure-b2-roomwise-r1-3 & 2 & F3 & 0.077 & 0.429 & 61.6 & 0.493 & 7 \\
pure-b3-roomwise-r3-3 & 3 & F3 & 0.081 & 0.433 & 61.8 & 0.515 & 9 \\
pure-b3-roomwise-r3-4 & 3 & F3 & 0.096 & 0.511 & 62.1 & 0.527 & 4 \\
\bottomrule
\end{tabular}

%% file: tableS6.tex
\begin{tabular}{rrrrrrrrrrr}
\toprule
$N$ & warned & runs & $I$ & $M$ & time & $\mathbb{E}_{\mathrm{mix}}[f_t]$ & $|\mathbb{E}_{\mathrm{mix}}[f_t]-\frac12|$ & $T_{\mathrm{lin}}$ & $I_{\mathrm{ind}}$ & $T_{\mathrm{ind}}$ \\
\midrule
20 & 0 & 3 & 0.229 & 0.697 & 71.1 & 0.695 & 0.195 & 66.1 & 0.198 & 67.8 \\
20 & 5 & 3 & 0.140 & 0.667 & 65.0 & 0.531 & 0.031 & 60.2 & 0.091 & 62.1 \\
20 & 10 & 3 & 0.200 & 0.292 & 68.6 & 0.367 & 0.133 & 62.8 & 0.145 & 64.7 \\
20 & 15 & 3 & 0.345 & 0.148 & 79.9 & 0.202 & 0.298 & 74.2 & 0.298 & 75.5 \\
20 & 20 & 3 & 0.468 & 0.064 & 95.4 & 0.038 & 0.462 & 94.1 & 0.462 & 94.4 \\
50 & 6 & 5 & 0.176 & 0.688 & 67.0 & 0.616 & 0.116 & 62.2 & 0.119 & 62.9 \\
50 & 12 & 5 & 0.127 & 0.651 & 63.5 & 0.537 & 0.037 & 60.2 & 0.064 & 61.0 \\
50 & 19 & 5 & 0.140 & 0.464 & 64.6 & 0.445 & 0.055 & 60.5 & 0.072 & 61.3 \\
50 & 25 & 5 & 0.202 & 0.291 & 67.3 & 0.367 & 0.133 & 62.8 & 0.135 & 63.6 \\
50 & 31 & 5 & 0.260 & 0.213 & 71.7 & 0.288 & 0.212 & 67.2 & 0.212 & 67.9 \\
50 & 38 & 5 & 0.336 & 0.141 & 78.4 & 0.196 & 0.304 & 74.8 & 0.304 & 75.3 \\
50 & 44 & 5 & 0.411 & 0.096 & 87.3 & 0.117 & 0.383 & 83.5 & 0.383 & 83.8 \\
\bottomrule
\end{tabular}

%% file: tableS8.tex
\begin{tabular}{rrrrrrrrrrrl}
\toprule
Block & $K$ & $I$ & $M$ & $M_{\mathrm{h}}$ & $M_{\mathrm{a}}$ & $\mathbb{E}[T_t]$ & $\mathbb{E}[T_{\mathrm{h},t}]$ & $\mathbb{E}[T_{\mathrm{a},t}]$ & $\mathbb{E}[f_{\mathrm{h},t}]$ & $\mathbb{E}[f_{\mathrm{a},t}]$ & frozen \\
\midrule
1 & 5 & 0.063 & 0.45 & 0.48 & 0.37 & 61.1 & 60.7 & 62.3 & 0.61 & 0.000 & no \\
2 & 5 & 0.063 & 0.50 & 0.50 & 0.51 & 61.2 & 61.3 & 60.9 & 0.66 & 0.000 & no \\
3 & 5 & 0.091 & 0.48 & 0.48 & 0.46 & 61.9 & 62.3 & 60.7 & 0.66 & 0.000 & no \\
4 & 5 & 0.059 & 0.45 & 0.46 & 0.40 & 60.9 & 60.9 & 61.0 & 0.67 & 0.000 & no \\
5 & 5 & 0.070 & 0.47 & 0.48 & 0.43 & 61.3 & 61.2 & 61.7 & 0.64 & 0.000 & no \\
6 & 5 & 0.079 & 0.49 & 0.49 & 0.49 & 61.3 & 61.3 & 61.5 & 0.64 & 0.000 & no \\
7 & 5 & \multicolumn{10}{l}{original attempt aborted after round 6; excluded from outcome summaries} \\
7$^{\mathrm{R}}$ & 5 & 0.069 & 0.44 & 0.43 & 0.49 & 61.1 & 61.2 & 60.9 & 0.64 & 0.000 & no \\
8 & 5 & \multicolumn{10}{l}{original attempt aborted after round 4; excluded from outcome summaries} \\
8$^{\mathrm{R}}$ & 5 & 0.061 & 0.49 & 0.52 & 0.43 & 61.3 & 61.6 & 60.3 & 0.68 & 0.000 & no \\
1 & 10 & 0.066 & 0.14 & 0.22 & 0.06 & 61.5 & 58.0 & 65.1 & 0.87 & 0.000 & no \\
2 & 10 & 0.084 & 0.20 & 0.30 & 0.09 & 61.7 & 57.0 & 66.5 & 0.85 & 0.010 & no \\
3 & 10 & 0.093 & 0.20 & 0.32 & 0.08 & 62.4 & 57.6 & 67.1 & 0.82 & 0.014 & no \\
4 & 10 & 0.063 & 0.17 & 0.26 & 0.07 & 61.1 & 57.6 & 64.6 & 0.88 & 0.012 & no \\
5 & 10 & 0.099 & 0.10 & 0.19 & 0.02 & 61.9 & 56.1 & 67.7 & 0.80 & 0.009 & no \\
6 & 10 & 0.107 & 0.22 & 0.31 & 0.13 & 62.6 & 57.1 & 68.1 & 0.78 & 0.019 & no \\
7 & 10 & 0.077 & 0.16 & 0.24 & 0.07 & 61.3 & 56.7 & 65.9 & 0.85 & 0.009 & no \\
8 & 10 & 0.094 & 0.20 & 0.31 & 0.09 & 62.2 & 57.4 & 67.0 & 0.81 & 0.023 & no \\
1 & 15 & 0.261 & 0.03 & 0.10 & 0.01 & 71.1 & 42.1 & 80.7 & 0.94 & 0.006 & yes \\
2 & 15 & 0.293 & 0.10 & 0.22 & 0.06 & 74.0 & 49.7 & 82.1 & 0.73 & 0.032 & yes \\
3 & 15 & 0.244 & 0.05 & 0.07 & 0.04 & 69.7 & 42.0 & 79.0 & 0.97 & 0.019 & no \\
4$^{\mathrm{F}}$ & 15 & 0.249 & 0.04 & 0.09 & 0.03 & 70.0 & 41.9 & 79.4 & 0.96 & 0.015 & no \\
5$^{\mathrm{F}}$ & 15 & 0.241 & 0.05 & 0.07 & 0.05 & 69.6 & 42.5 & 78.6 & 0.97 & 0.023 & no \\
6$^{\mathrm{R}}$ & 15 & 0.259 & 0.06 & 0.15 & 0.03 & 70.9 & 43.4 & 80.1 & 0.91 & 0.017 & yes \\
7 & 15 & 0.269 & 0.07 & 0.21 & 0.02 & 72.0 & 44.4 & 81.2 & 0.89 & 0.011 & yes \\
8$^{\mathrm{R}}$ & 15 & 0.260 & 0.03 & 0.10 & 0.01 & 71.0 & 42.0 & 80.6 & 0.94 & 0.006 & yes \\
\bottomrule
\end{tabular}
\par\smallskip{\footnotesize Unmarked rows: initial collection, 14--15 September 2026. $^{\mathrm{R}}$: 16 September; $^{\mathrm{F}}$: 17 September. The original aborted attempts are retained as separate rows.}

%% file: tableS9.tex
\begin{tabular}{llrrrrrrrrrr}
\toprule
Arm & seed & $T_{30}$ & $f_{31}$ & $I_{31}$ & $M_{31}$ & $T_{31}$ & $T_{32}$ & $T_{33}$ & $T_{34}$ & $T_{35}$ & $\mathbb{E}_{36:50}[T_t]$ \\
\midrule
F3$\rightarrow$F1 & 3 & 91.0 & 0.96 & 0.46 & 0.94 & 93.9 & 60.3 & 70.8 & 80.7 & 60.6 & 72.3 \\
F3$\rightarrow$F1 & 4 & 80.7 & 1.00 & 0.50 & 0.86 & 100.0 & 63.1 & 61.6 & 69.2 & 64.1 & 71.3 \\
F3$\rightarrow$F1 & 5 & 93.9 & 0.96 & 0.46 & 0.92 & 93.9 & 60.1 & 70.8 & 83.1 & 60.6 & 73.6 \\
F3$\rightarrow$F1 & 6 & 88.2 & 0.98 & 0.48 & 0.90 & 96.9 & 70.8 & 60.3 & 76.4 & 61.0 & 73.1 \\
F3$\rightarrow$F1 & 7 & 96.9 & 0.96 & 0.46 & 0.98 & 93.9 & 64.1 & 67.7 & 83.1 & 63.1 & 67.2 \\
F1$\rightarrow$F3 & 3 & 91.0 & 0.00 & 0.50 & 0.06 & 100.0 & 100.0 & 93.9 & 93.9 & 96.9 & 95.4 \\
F1$\rightarrow$F3 & 4 & 60.1 & 0.00 & 0.50 & 0.48 & 100.0 & 100.0 & 96.9 & 96.9 & 93.9 & 90.8 \\
F1$\rightarrow$F3 & 5 & 70.8 & 0.00 & 0.50 & 0.24 & 100.0 & 100.0 & 100.0 & 96.9 & 100.0 & 93.3 \\
F1$\rightarrow$F3 & 6 & 60.3 & 0.00 & 0.50 & 0.46 & 100.0 & 96.9 & 88.2 & 100.0 & 100.0 & 96.0 \\
F1$\rightarrow$F3 & 7 & 72.5 & 0.00 & 0.50 & 0.22 & 100.0 & 100.0 & 100.0 & 96.9 & 100.0 & 92.5 \\
\bottomrule
\end{tabular}

%% file: tableS10.tex
\begin{tabular}{rrrrr}
\toprule
Block & $b_b(I)$ & $b_b(M)$ & LOO mean $I$ slope & LOO mean $M$ slope \\
\midrule
1 & +0.0993 & -0.2086 & +0.0945 & -0.2080 \\
2 & +0.1150 & -0.2007 & +0.0922 & -0.2091 \\
3 & +0.0764 & -0.2150 & +0.0978 & -0.2070 \\
4 & +0.0950 & -0.2007 & +0.0951 & -0.2091 \\
5 & +0.0857 & -0.2079 & +0.0964 & -0.2081 \\
6 & +0.0900 & -0.2129 & +0.0958 & -0.2073 \\
7 & +0.1000 & -0.1886 & +0.0944 & -0.2108 \\
8 & +0.0993 & -0.2300 & +0.0945 & -0.2049 \\
\bottomrule
\end{tabular}

%% file: tableS11.tex
\begin{tabular}{lrrr}
\toprule
Nuisance variable & $\gamma$ & Mean $I$ slope & Mean $M$ slope \\
\midrule
Carryover & -0.050 & +0.0982 & -0.2049 \\
Carryover & -0.020 & +0.0963 & -0.2068 \\
Carryover & -0.010 & +0.0957 & -0.2074 \\
Carryover & -0.005 & +0.0954 & -0.2077 \\
Carryover & +0.000 & +0.0951 & -0.2080 \\
Carryover & +0.005 & +0.0948 & -0.2083 \\
Carryover & +0.010 & +0.0945 & -0.2087 \\
Carryover & +0.020 & +0.0938 & -0.2093 \\
Carryover & +0.050 & +0.0920 & -0.2112 \\
Wait (min) & -0.005 & +0.0957 & -0.2074 \\
Wait (min) & +0.000 & +0.0951 & -0.2080 \\
Wait (min) & +0.005 & +0.0944 & -0.2087 \\
\bottomrule
\end{tabular}

%% file: tableS12.tex
\begin{tabular}{rrrrrrrr}
\toprule
$K$ & Actual (\%) & Responses & Mean (\%) & SD & Median [IQR] & Bias (pp) & MAE (pp) \\
\midrule
5 & 25 & 116/120 & 63.1 & 24.9 & 65.0 [50.0, 80.0] & +38.1 & 39.4 \\
10 & 50 & 76/80 & 67.2 & 19.0 & 70.0 [50.0, 80.0] & +17.2 & 20.4 \\
15 & 75 & 38/40 & 68.9 & 21.2 & 70.0 [60.0, 83.0] & -6.1 & 17.1 \\
\bottomrule
\end{tabular}

%% file: tableS13.tex
\begin{tabular}{lrrrrrrr}
\toprule
Message & frozen & $I$ & $M$ & time (min) & $\mathbb{E}[f_t]$ & response & variation \\
\midrule
\multicolumn{8}{l}{\emph{GPT-5.4-mini, no reasoning, temperature 0.7 (core runs at the same seeds)}} \\
tip & 0/5 & 0.146 & 0.714 & 65.2 & 0.555 & 0.7 & 4.6 \\
tip + warning & 5/5 & 0.460 & 0.079 & 94.0 & 0.040 & 33.9 & 0.2 \\
warning $-$ tip & & +0.314 (5) & $-$0.635 (5) & +28.8 (5) & $-$0.515 (5) & +33.2 & $-$4.4 \\
\addlinespace
\multicolumn{8}{l}{\emph{GPT-5.4-mini, no reasoning, temperature 1.0}} \\
tip & 0/5 & 0.131 & 0.687 & 63.9 & 0.495 & 0.1 & 3.7 \\
tip + warning & 5/5 & 0.442 & 0.115 & 91.5 & 0.058 & 31.3 & 0.2 \\
warning $-$ tip & & +0.311 (5) & $-$0.572 (5) & +27.6 (5) & $-$0.437 (5) & +31.2 & $-$3.4 \\
\addlinespace
\multicolumn{8}{l}{\emph{GPT-5.4-mini, low reasoning effort (provider default temperature)}} \\
tip & 0/5 & 0.189 & 0.655 & 66.9 & 0.690 & 5.8 & 1.2 \\
tip + warning & 0/5 & 0.241 & 0.416 & 70.0 & 0.259 & 9.3 & 0.8 \\
warning $-$ tip & & +0.051 (5) & $-$0.240 (5) & +3.2 (5) & $-$0.430 (5) & +3.5 & $-$0.5 \\
\addlinespace
\multicolumn{8}{l}{\emph{GPT-5.4-mini, medium reasoning effort (provider default temperature)}} \\
tip & 0/5 & 0.097 & 0.582 & 62.3 & 0.582 & 1.1 & 1.3 \\
tip + warning & 0/5 & 0.295 & 0.360 & 74.7 & 0.205 & 13.9 & 0.9 \\
warning $-$ tip & & +0.198 (5) & $-$0.222 (5) & +12.4 (5) & $-$0.377 (5) & +12.8 & $-$0.5 \\
\addlinespace
\multicolumn{8}{l}{\emph{GPT-6 Luna, no reasoning (no temperature sent)}} \\
tip & 0/5 & 0.184 & 0.626 & 67.8 & 0.355 & 3.4 & 4.1 \\
tip + warning & 0/5 & 0.323 & 0.357 & 78.4 & 0.185 & 15.9 & 2.5 \\
warning $-$ tip & & +0.139 (5) & $-$0.269 (5) & +10.6 (5) & $-$0.170 (5) & +12.5 & $-$1.6 \\
\addlinespace
\multicolumn{8}{l}{\emph{GPT-6 Luna, medium reasoning effort, the model default (no temperature sent)}} \\
tip & 5/5 & 0.448 & 0.103 & 92.4 & 0.052 & 32.2 & 0.2 \\
tip + warning & 5/5 & 0.427 & 0.145 & 89.5 & 0.073 & 29.2 & 0.3 \\
warning $-$ tip & & $-$0.022 (1) & +0.042 (0) & $-$2.9 (2) & +0.022 (1) & $-$3.0 & +0.1 \\
\bottomrule
\end{tabular}

%% file: tableS14.tex
\begin{tabular}{lrrrrrrl}
\toprule
Message & seed & $I$ & $M$ & time (min) & $\mathbb{E}[f_t]$ & ties & state \\
\midrule
\multicolumn{8}{l}{\emph{GPT-5.4-mini, no reasoning, temperature 1.0}} \\
tip & 3 & 0.139 & 0.691 & 64.3 & 0.519 & 3 & near equilibrium \\
tip & 4 & 0.126 & 0.686 & 63.6 & 0.467 & 2 & near equilibrium \\
tip & 5 & 0.133 & 0.681 & 64.0 & 0.519 & 1 & near equilibrium \\
tip & 6 & 0.127 & 0.692 & 64.0 & 0.494 & 5 & near equilibrium \\
tip & 7 & 0.131 & 0.688 & 63.9 & 0.477 & 2 & near equilibrium \\
tip + warning & 3 & 0.451 & 0.098 & 92.7 & 0.049 & 0 & frozen \\
tip + warning & 4 & 0.431 & 0.137 & 89.9 & 0.069 & 0 & frozen \\
tip + warning & 5 & 0.454 & 0.092 & 93.1 & 0.046 & 0 & frozen \\
tip + warning & 6 & 0.430 & 0.139 & 89.9 & 0.070 & 0 & frozen \\
tip + warning & 7 & 0.445 & 0.109 & 91.9 & 0.055 & 0 & frozen \\
\addlinespace
\multicolumn{8}{l}{\emph{GPT-5.4-mini, low reasoning effort (provider default temperature)}} \\
tip & 3 & 0.185 & 0.651 & 66.7 & 0.686 & 2 & partial \\
tip & 4 & 0.200 & 0.659 & 67.4 & 0.700 & 1 & partial \\
tip & 5 & 0.186 & 0.651 & 66.8 & 0.690 & 2 & partial \\
tip & 6 & 0.185 & 0.653 & 66.6 & 0.687 & 1 & partial \\
tip & 7 & 0.190 & 0.662 & 66.9 & 0.686 & 0 & partial \\
tip + warning & 3 & 0.240 & 0.419 & 70.1 & 0.260 & 0 & partial \\
tip + warning & 4 & 0.251 & 0.404 & 70.6 & 0.249 & 0 & partial \\
tip + warning & 5 & 0.227 & 0.420 & 68.9 & 0.273 & 0 & partial \\
tip + warning & 6 & 0.255 & 0.407 & 71.1 & 0.245 & 0 & partial \\
tip + warning & 7 & 0.232 & 0.429 & 69.5 & 0.268 & 0 & partial \\
\addlinespace
\multicolumn{8}{l}{\emph{GPT-5.4-mini, medium reasoning effort (provider default temperature)}} \\
tip & 3 & 0.108 & 0.588 & 63.4 & 0.604 & 5 & near equilibrium \\
tip & 4 & 0.107 & 0.586 & 62.6 & 0.588 & 2 & near equilibrium \\
tip & 5 & 0.085 & 0.579 & 61.7 & 0.567 & 3 & near equilibrium \\
tip & 6 & 0.094 & 0.572 & 62.1 & 0.578 & 4 & near equilibrium \\
tip & 7 & 0.091 & 0.583 & 61.9 & 0.573 & 3 & near equilibrium \\
tip + warning & 3 & 0.291 & 0.371 & 74.4 & 0.209 & 0 & partial \\
tip + warning & 4 & 0.299 & 0.346 & 75.2 & 0.201 & 0 & partial \\
tip + warning & 5 & 0.287 & 0.375 & 73.9 & 0.213 & 0 & partial \\
tip + warning & 6 & 0.299 & 0.349 & 75.1 & 0.201 & 0 & partial \\
tip + warning & 7 & 0.297 & 0.357 & 75.2 & 0.203 & 0 & partial \\
\addlinespace
\multicolumn{8}{l}{\emph{GPT-6 Luna, no reasoning (no temperature sent)}} \\
tip & 3 & 0.177 & 0.629 & 67.2 & 0.354 & 3 & partial \\
tip & 4 & 0.170 & 0.650 & 66.4 & 0.357 & 0 & partial \\
tip & 5 & 0.192 & 0.621 & 68.4 & 0.353 & 2 & partial \\
tip & 6 & 0.199 & 0.610 & 69.2 & 0.355 & 2 & partial \\
tip & 7 & 0.181 & 0.620 & 67.7 & 0.355 & 3 & partial \\
tip + warning & 3 & 0.328 & 0.361 & 79.8 & 0.198 & 0 & partial \\
tip + warning & 4 & 0.313 & 0.372 & 77.7 & 0.187 & 0 & partial \\
tip + warning & 5 & 0.339 & 0.330 & 80.2 & 0.174 & 0 & partial \\
tip + warning & 6 & 0.316 & 0.363 & 77.5 & 0.184 & 0 & partial \\
tip + warning & 7 & 0.318 & 0.359 & 76.9 & 0.182 & 0 & partial \\
\addlinespace
\multicolumn{8}{l}{\emph{GPT-6 Luna, medium reasoning effort, the model default (no temperature sent)}} \\
tip & 3 & 0.456 & 0.088 & 93.4 & 0.044 & 0 & frozen \\
tip & 4 & 0.436 & 0.128 & 90.5 & 0.064 & 0 & frozen \\
tip & 5 & 0.463 & 0.074 & 94.4 & 0.037 & 0 & frozen \\
tip & 6 & 0.435 & 0.129 & 90.5 & 0.065 & 0 & frozen \\
tip & 7 & 0.453 & 0.094 & 93.0 & 0.047 & 0 & frozen \\
tip + warning & 3 & 0.420 & 0.157 & 88.6 & 0.080 & 0 & frozen \\
tip + warning & 4 & 0.436 & 0.129 & 90.6 & 0.064 & 0 & frozen \\
tip + warning & 5 & 0.419 & 0.161 & 88.5 & 0.081 & 0 & frozen \\
tip + warning & 6 & 0.435 & 0.129 & 90.5 & 0.065 & 0 & frozen \\
tip + warning & 7 & 0.424 & 0.148 & 89.2 & 0.076 & 0 & frozen \\
\bottomrule
\end{tabular}